\documentclass[twocolumn,epjc3]{svjour3}          

\RequirePackage[T1]{fontenc}

\RequirePackage{graphicx}
\RequirePackage{mathptmx}      
\RequirePackage[numbers,sort&compress]{natbib}
\RequirePackage[colorlinks,citecolor=blue,urlcolor=blue,linkcolor=blue]{hyperref}

\journalname{Eur. Phys. J. C}

\usepackage{epsfig,wrapfig}
\usepackage{bm}
\usepackage{amssymb,amsmath,slashed,amsfonts,latexsym}
\usepackage[normalem]{ulem} 
\renewcommand\sout{\bgroup \color[rgb]{1,0.55,0} \ULdepth=-.5ex \ULset}

\usepackage{dsfont}
\usepackage{slashbox}

\newcommand{\be}{\begin{equation}}
\newcommand{\ee}{\end{equation}}
\newcommand{\ba}{\begin{eqnarray}}
\newcommand{\ea}{\end{eqnarray}}
\newcommand{\la}{\langle}
\newcommand{\ra}{\rangle}
\newcommand{\di}{\mathrm{d}}
\newcommand{\ud}{\mathrm{d}}

\newcommand{\uTr}{\mathrm{Tr}}

\newcommand{\uslash}{/\!\!\!}

\newlength\savedwidth

\newcommand{\uvec}[1]{\bm{#1}}

\def\ket#1{\hbox{$\vert #1\rangle$}}   

\allowdisplaybreaks[2]

\begin{document}

\title{Transverse pion structure beyond leading twist in constituent models
}


\author{C.~Lorc\'e 
        \and
       B. Pasquini 
       \and
       P. Schweitzer 
}



\institute{Centre de Physique Th\'eorique, \'Ecole polytechnique, 
	CNRS, Universit\'e Paris-Saclay, F-91128 Palaiseau, France \label{addr1}
	\and
        Dipartimento di Fisica, 
  	Universit\`a degli Studi di Pavia, Pavia, Italy \label{addr2}
        \and
        Istituto Nazionale di Fisica Nucleare, 
 	 Sezione di Pavia, Pavia, Italy\label{addr3}
  	\and 
  	Department of Physics, University of Connecticut, 
  	Storrs, CT 06269, USA\label{addr4}
  	\and
  	Institute for Theoretical Physics, T\"ubingen University, 
	Auf der Morgenstelle 14, 72076 T\"ubingen, Germany\label{addr5}
}

\date{Received: date / Accepted: date}

\maketitle

\begin{abstract}
The understanding of the pion structure as described in terms of
transverse-momentum dependent parton distribution functions (TMDs)
is of importance for the interpretation of currently ongoing 
Drell-Yan experiments with pion beams. 
In this work we discuss the description of pion TMDs beyond leading twist 
in a pion model formulated in the light-front constituent framework. 
For comparison, we also review and derive new results for pion TMDs in 
the bag and spectator model.
          
\end{abstract}
\section{Introduction}
\label{Sec-1:introduction}

The pion is one of the few hadrons, besides nucleon and nuclei, whose 
partonic structure can be studied, mainly thanks to the Drell-Yan process (DY)
\cite{Drell:1970wh,Christenson:1970um} with pion beams impinging on nuclear
targets \cite{Peng:2014hta,Chang:2013opa,Reimer:2007iy,McGaughey:1999mq}.
DY data provide access to the twist-2 ``collinear'' parton distribution 
function (PDF) of the pion $f_1^a(x)$ 
\cite{Owens:1984zj,Gluck:1991ey,Sutton:1991ay,Gluck:1999xe,Hecht:2000xa,
Wijesooriya:2005ir,Holt:2010vj,Aicher:2010cb} and more.
In fact, the unpolarized DY cross section differential in the dilepton angular
distribution, given in the Collins-Soper frame \cite{Collins:1981uw} by 
\be\label{Eq:DY-unp-angular-dependence}
\frac{\ud\sigma}{\ud \Omega} \propto 
 \bigg( 1 + \lambda \cos^2 \theta
      + \mu \sin 2\theta \cos \phi
      + \frac{\nu}{2} \sin^2 \theta \cos 2\phi \bigg) \; ,
\ee
provides also information on transverse momentum dependent parton
distribution functions (TMDs). In the TMD factorization framework,
the coefficient $\lambda$ is due to the twist-2 unpolarized TMD $f_1^q(x,p_T)$
and $1/Q^2$-suppressed terms, $\mu$ arises from certain twist-3 TMDs 
\cite{Arnold:2008kf}, $\nu$ is due to the naive time-reversal 
odd ($\mathsf T$-odd) Boer-Mulders function \cite{Boer:1999mm}. 
One important current development consists in extending the DY measurements 
to include polarization effects, which is being pursued with polarized proton 
beams at RHIC (BNL) \cite{Aschenauer:2015eha} and pion beams impinging on 
polarized proton targets at COMPASS (CERN) \cite{Quintans:2011zz,Gautheron:2010wva}. 
These experiments will test the TMD factorization approach, in
particular the predicted sign change of naive $\mathsf T$-odd
TMDs \cite{Collins:2002kn}, and provide new insights on the nucleon structure.

In our context, the COMPASS program is of particular interest.
It will give at the same time new insights on the pion structure
at leading {\it and} subleading twist, and will go far beyond what 
was learned from earlier Fermilab and CERN experiments 
\cite{Falciano:1986wk,Guanziroli:1987rp,Conway:1989fs} owing
to the availability of a polarized target. Moreover, previous 
measurements suffered from limited statistics, and most of them found
for instance a subleading-twist coefficient $\mu$ compatible with zero. 
Also with this respect new data from COMPASS may improve the situation 
\cite{Quintans:2011zz}.

Higher-twist PDFs and TMDs are of interest in their own right, 
as they provide a window on quark-gluon dynamics. By exploring the 
equations of motion (EOM) of QCD, higher-twist PDFs and TMDs can in 
general be decomposed into contributions from leading-twist, current quark
mass terms and pure quark-gluon interaction-dependent (``tilde'') terms. 
An interesting question is how such genuine QCD interaction-dependent 
terms are modeled in constituent frameworks, which for our purposes 
are defined as models without explicit gluon degrees of freedom.

In a previous study we addressed this question in the context of
unpolarized nucleon PDFs and TMDs \cite{Lorce:2014hxa}.
We have shown that
internally consistent descriptions of the unpolarized leading- and
higher-twist PDFs and TMDs are possible using several constituent model approaches. The respective effective
interactions mimic in various ways the QCD quark-gluon interactions,
giving rise to non-trivial 
tilde-terms in some models.
To which extent constituent models can provide phenomenologically
reliable estimates for higher-twist effects remains to be tested.
At least an encouraging agreement was observed 
\cite{Lorce:2014hxa} in the case of the nucleon twist-3 
PDF $e^q(x)$ of which recently a first extraction became available 
\cite{Courtoy:2014ixa}.

In this work we will present a study for the pion case.
The main scope is to prepare an understanding of $\mathsf T$-even pion TMDs at 
leading and especially subleading twist in the framework of constituent 
models which can be tested and used in future phenomenological applications 
to analyze and interpret first data. Our particular focus will be on 
critically reviewing the internal consistency of the models, and assess 
their range of applicability. We will also investigate how the genuine 
higher-twist terms are modeled in different effective-model frameworks.
Our focus will be on the aspects peculiar to the meson sector, 
i.e.\ on aspects related to the modeling of 2-body dynamics of 
the $q\bar q$-pair in the pion as opposed to the modeling
of 3-body dynamics in the nucleon state 
investigated in prior work \cite{Lorce:2014hxa}.

The three models discussed in this work are the light-front constituent 
model (LFCM), bag and spectator model. 
All results for higher-twist TMDs are new and original in the LFCM
and bag model. In the spectator model analytical expressions for
twist-3 pion TMDs were quoted in literature, but to the best of our knowledge
they were neither evaluated nor were their properties discussed.
We discuss and compare the results from the different models 
with the goal to establish differences and
common features of constituent frameworks of the pion structure.

It is important to keep in mind that none of these models accounts for the 
perhaps most important feature of the pion, namely its nature as Goldstone 
boson associated with spontaneous chiral symmetry breaking. 
Instead, the models discussed in this work treat the pion on the same footing 
as all other hadrons, i.e.\ as a particle composed of the respective constituent
degrees of freedom. In our assessment of the applicability of the models, 
we shall also discuss the rational for this approach.
A study of twist-2 pion TMDs in a chiral (Nambu-Jona-Lasinio) 
model was presented in Ref.~\cite{Noguera:2015iia}.

The outline is as follows. In section~\ref{Sec-2:T-even-TMDs-of-pion}
we define and discuss the properties of pion TMDs in constituent models.
In section~\ref{Sec-3:LFCM} we study pion TMDs in the LFCM. In
section~\ref{Sec-4:other-models} we review the descriptions of pion 
TMDs in the bag and spectator model.
In section~\ref{Sec-4:results} we present the numerical results 
from the different models and compare them to nucleon TMDs. 
Finally, section~\ref{Sec-5:conclusions} contains the conclusions.
Technical details are collected in the appendices.

\section{$\mathsf T$-even pion TMD\lowercase{s} in quark models}
\label{Sec-2:T-even-TMDs-of-pion}

TMDs are described in terms of quark correlators. 
In constituent approaches without explicit gluon degrees of freedom, 
the Wilson lines of QCD reduce to unit matrices in color space. As a result 
$\mathsf T$-odd TMDs are absent, and only $\mathsf T$-even TMDs appear.
The structure of a spin-zero hadron, like the pion, is described 
in terms of 4 TMDs,
\begin{subequations}
   \label{Eq:correlator-TMDs0}
\begin{align}
   \label{Eq:correlator-TMDs1}
   \int \frac{\di z^-\di^2z_T}{2(2\pi)^3} \, 
   e^{i p \cdot z} \, \la P|\overline{\psi}(0)\gamma^+ \psi(z)|P\ra|_{z^+=0} 
   &= f_1^q(x,p_T),
   \\
   \label{Eq:correlator-TMDs2}
   \int \frac{\di z^-\di^2z_T}{2(2\pi)^3} e^{i p \cdot z} \la P  |  
   \overline{\psi}(0)\;\mathds{1}\; \psi(z)  |  P\ra|_{z^+=0} 
   &= \frac{m_\pi}{P^+}\,e^q(x,p_T),
   \\
   \label{Eq:correlator-TMDs3}
   \int \frac{\di z^-\di^2z_T}{2(2\pi)^3} e^{i p \cdot z} \la P |  
   \overline{\psi}(0)\gamma_T^j \psi(z)  |  P\ra|_{z^+=0} 
   &= \frac{p_T^j}{P^+}\,f^{\perp q}(x,p_T),
   \\
   \label{Eq:correlator-TMDs4}
   \int \frac{\di z^-\di^2z_T}{2(2\pi)^3} e^{i p \cdot z} \la P  |  
   \overline{\psi}(0)\gamma^- \psi(z)  |  P\ra|_{z^+=0} 
   &= \frac{m_\pi^2}{(P^+)^2}f_4^q(x,p_T).
\end{align}
\end{subequations}
Here $|P\ra$ is a pion state with 4-momentum $P$, $q$ is a flavor 
index for the quark and antiquark  contribution and $m_\pi$ is the 
pion mass. We use light-front coordinates $a^\pm = (a^0 \pm a^3)/\sqrt{2}$,
$\bm{a}_T=(a^1,a^2)$ with $a_T\equiv|\bm{a}_T|$ and the metric is
$a\cdot b$ = $a^+b^-+a^-b^+-\bm{a}_T\cdot\bm{b}_T$
with $\di^4 z = \di z^+\di z^-\di^2 z_T$. 
The model results generically refer to a low (``hadronic'') normalization 
scale below 1 GeV \cite{Boffi:2009sh,Pasquini:2011tk,Pasquini:2014ppa}. 
Integrating Eq. \eqref{Eq:correlator-TMDs0} over $\uvec p_T$ provides 
the definition of the corresponding PDFs. Note in particular that because 
of the explicit $p^j_T$ factor in Eq. \eqref{Eq:correlator-TMDs3} there does 
not exist any PDF counterpart to $f^{\perp q}(x,p_T)$. One can however formally 
define $f^{\perp q}(x)\equiv\int\ud^2p_T\,f^{\perp q}(x,p_T)$.

Sum rules are of particular importance when testing the consistency of models.
Let $N_q$ be the valence number of flavor $q$, which is for instance
$N_u=N_{\bar d}=1$ in $\pi^+$. The sum rules are given by
\begin{subequations}\begin{align}
  \int\di x\,f_1^q(x) 	& = N_q\,, 	\label{Eq:f1-sum-rule} \\
  \sum_q\int\di x\;x\,f_1^q(x)& = 1  \,, \label{Eq:mom-sum-rule} \\
  \sum_q\int\di x\,e^q(x)& = \frac{\sigma_{\pi}}{m_q}\,,
					\label{Eq:sigma-sum-rule} \\
  \int\di x\;x\;e^q(x) 	& = \frac{m_q}{m_\pi}\;N_q\,,
					\label{Eq:Jaffe-Ji-sum-rule} \\
  2\int\di x\,f_4^q(x) 	& = N_q. 	\label{Eq:f4-sum-rule}
\end{align}\end{subequations}
The valence number sum rule (\ref{Eq:f1-sum-rule}) is the same in QCD 
and constituent models, but the momentum sum rule \eqref{Eq:mom-sum-rule}
is saturated solely by valence degrees of freedom in constituent models 
at the initial scale (with the exception of the spectator model which 
we will discuss in detail).
Equation~(\ref{Eq:sigma-sum-rule}) formally relates $e^q(x)$ to the 
sigma term \cite{Jaffe:1991ra,Efremov:2002qh}, which corresponds 
to the scalar form factor $\sigma(t)$ at zero-momentum transfer. 
The sigma term of the pion is given by 
$\sigma_\pi = \frac12\,m_\pi$ in the leading order of the chiral expansion.
Since $m_\pi^2\propto m_q$ owing to the Gell-Mann--Oakes--Renner relation,
the sum rule (\ref{Eq:sigma-sum-rule}) for the pion diverges like $1/m_\pi$ 
in the chiral limit.
The Jaffe-Ji sum rule (\ref{Eq:Jaffe-Ji-sum-rule}) connects the first moment 
of $e^q(x)$ to the current quark mass $m_q$ in QCD, or the constituent (or 
effective) mass in models \cite{Lorce:2014hxa,Schweitzer:2003uy}.
In the chiral limit this sum rule goes to zero like $m_\pi$.
The sum rule (\ref{Eq:f4-sum-rule}) formally arises from the normalization 
of the minus-component of the vector current, just as (\ref{Eq:f1-sum-rule})
arises from the normalization of its plus-component. The validity 
of (\ref{Eq:f4-sum-rule}) is subtle, both in QCD and in quark models 
\cite{Lorce:2014hxa}, as we shall discuss in sections~\ref{Sec-3:LFCM}
and \ref{Sec-4:other-models}.
In equation~(\ref{Eq:sigma-sum-rule}) and throughout this work,
we neglect isospin-violating effects and assume $m_q=m_u=m_d$ for current 
or constituent quark masses. Unless otherwise stated, we will refer to the 
distributions in positive pions using the notation 
$j^{u}_{\pi^+}(x,p_T)\equiv j^q(x,p_T)$, where 
$j_{\pi^{+}}^{u}=j_{\pi^{+}}^{\bar d}
=j_{\pi^{-}}^{d}=j_{\pi^{-}}^{\bar u}
=2 \,j_{\pi^{0}}^{u} = 2 \,j_{\pi^{0}}^{\bar u} 
=2 \,j_{\pi^{0}}^{d} = 2 \,j_{\pi^{0}}^{\bar d}$ 
holds due to isospin symmetry and charge conjugation, and $j^q(x,p_T)$ denotes a generic TMD. 

Positivity inequalities provide another important test, although they can be 
spoiled in QCD already at leading twist (let alone at twist-4) due to 
subtractions in the renormalization procedure. In consistent models one 
expects \cite{Lorce:2014hxa}
\begin{subequations}
\begin{align}
	f_1^q(x,p_T) \ge 0, \label{Eq:f1-inequality} & \\
    	f_4^q(x,p_T) \ge 0. \label{Eq:f4-inequality} &
\end{align}
\end{subequations}

In approaches
without explicit gauge-degrees of freedom,
the quark correlator of a spin-zero (or unpolarized) hadron has a general 
Lorentz decomposition in 
terms of 3 independent amplitudes parametrized in terms of 4 TMDs. 
In such situations ``quark-model Lorentz-invariance relations (qLIRs)''
arise \cite{Tangerman:1994bb}\footnote{We stress 
	that such relations are valid only in quark models
        (or for the (academic) TMDs with straight gauge links
        implemented in lattice calculations \cite{Hagler:2009mb}),
        and should be distinguished from the LIRs in collinear twist-3
        formalism recently derived in Ref.~\cite{Kanazawa:2015ajw}.}.
In our case, the qLIR is given by \cite{Lorce:2014hxa}
\be
    f_4^q(x) = \frac{1}{2}\,f_1^q(x)+\frac{\di}{\di x}f^{\perp q(1)}(x),
    \label{Eq:LIR-f4}
\ee
with $f^{\perp q(1)}(x)=\int\di^2p_T\,\tfrac{p^2_T}{2m_\pi^2}f^{\perp q}(x,p_T)$.

It is important to remark that $f_1^q$ and the twist-3 pion TMDs $e^q$ 
and $f^{\perp q}$ can be accessed in DY \cite{Arnold:2008kf}, but not the 
twist-4 TMD $f^q_4$ which therefore has to be considered as an academic object.
Nevertheless $f^q_4$ completes the description of the quark correlator 
through twist-4 \cite{Goeke:2005hb}, and the relation (\ref{Eq:LIR-f4}) 
is of value as it provides a powerful test for the theoretical consistency 
of a model.

Next, let us state the relations which result from employing the EOMs
\begin{subequations}
\begin{align}
	x\,e^q(x,p_T) & = 
	x\,\tilde{e}^q(x,p_T) + \frac{m_q}{m_\pi}\,f_1^q(x,p_T),\label{Eq:eom-e}\\
 	x\,f^{\perp q}(x,p_T) & = 
	x\,\tilde{f}^{\perp q}(x,p_T) + f_1^q(x,p_T),\label{Eq:eom-fperp}\\
	x^2f^q_4(x,p_T) & = 
	x^2\tilde{f}_4^q(x,p_T) + \frac{p_T^2+m_q^2}{2m_\pi^{2}}\;f_1^q(x,p_T).
	\label{Eq:eom-f4}
\end{align}
\end{subequations}
In QCD the tilde-terms are expressed in terms of quark-gluon-quark
correlators. In quark models, they still denote ``interaction-dependent
terms'' which arise from applying the respective model EOMs.

\section{Pion structure in the LFCM}
\label{Sec-3:LFCM}

In this section we discuss pion TMDs 
in the LFCM.
We first derive the general expressions for the subleading-twist TMDs 
in leading order of the Fock space expansion for the pion, and discuss 
the consistency of the approach. We then introduce the phenomenological 
model for the light-front wave-functions (LFWFs) which we will employ 
later to obtain definite predictions.

\subsection{General formalism}
\label{Sec-3a:LFCM}

The formalism for the calculation of the unpolarized higher-twist 
$\mathsf T$-even TMDs 
in the light-front framework has been discussed in Ref.~\cite{Lorce:2014hxa}, 
with an explicit application to the nucleon.
The same approach is adopted here in the case of pion. We recall that in light-front quantization the Fock-space expansion of the hadron states  is performed
in terms of free on-mass-shell parton states with the essential QCD bound-state information encoded in the LFWF.
The $q\bar q$ component of the light-front state of the pion can be written as 
\be\label{LFWF}
  \ket{\pi(\tilde P)}_{q\bar q} = \sum_{\lambda_1,\lambda_2}
  \int \ud[1]\,\ud[2]\,   \Psi^{q \bar q}_{\lambda_1\lambda_2}(r_1,r_2)      
  |\lambda_i, \tilde p_i \rangle\, ,
\ee
where $\Psi^{q \, \bar q}_{\lambda_1\lambda_2}$ is the $q\bar q$-LFWF with 
$\lambda_1$  ($\lambda_2$) and $q$ ($\bar q$) referring to the 
light-front helicity and flavor of quark (antiquark), respectively.
The LFWF includes an isospin factor $T_\pi$ which projects
onto the different members of the isotriplet of the pion, i.e.\
$T_\pi=\sum_{\tau_1,\tau_{2}}\langle \tfrac{1}{2} \tau_1 \tfrac{1}{2} \tau_{2}| 1 \tau_\pi \rangle$ 
with $\tau_1, \tau_{2}$ and $\tau_\pi$ the isospin of the quark, antiquark
and pion state, respectively. In equation~\eqref{LFWF}
$r_i=(x_i M_{0} , \bm{p}_{Ti})$,
and $M_0$ denotes the mass of the non-interacting $q\bar q$ state. 
Furthermore, we introduced the notation 
$\tilde p=(p^+,\bm{p}_T)$ for a generic light-front momentum 
variable $p$.
Since momentum conservation implies $\bm p_{T1}+\bm p_{T2}=\bm{0}_T$ 
and $x_1+x_2=1$,  the LFWF actually depends only on the variables $\bar x=x_1$ and  
$\bm{\kappa}_{T}=\bm p_{T1}$. The integration measure in 
Eq.~(\ref{LFWF}) is defined as
\begin{eqnarray}
   \ud[1]\,\ud[2]&=&
   \frac{\ud x_1\,\ud x_2 }{\sqrt{x_1x_2}}\,\delta\!\left(1-x_1-x_2\right)\nonumber\\
  && \times \frac{\ud^2 p_{T1}\,\ud^2p_{T2}}{2(2\pi)^3} \,
   \delta^{(2)}\!\left(\bm p_{T1}+\bm p_{T2}\right),
   \label{eq:7}
   \end{eqnarray}
so that we can write
\begin{eqnarray}
	&&\int \ud[1]\,\ud[2]\,F(x_1,\bm{p}_{T1},x_2,\bm{p}_{T2})
	\nonumber\\
	&&=\int \frac{\ud\bar x}{\sqrt{\bar x(1-\bar x)}}   
   	\,\frac{\ud^2\kappa_{T}}{2(2\pi)^3}\,
	F( \bar x,\bm{\kappa}_T, 1-\bar x, -\bm{\kappa}_T).
\end{eqnarray}

The pion TMDs are given by the expressions
\begin{subequations}\begin{align}
   f_1^q(x,p_T)&= {\cal P}^q(\tilde p),\label{f1:lfwf}\\
   x\,e^q(x,p_T)&=\frac{m_q}{m_\pi}\, {\cal P}^q(\tilde p),\label{e:lfwf}\\
   x\,f^{\perp q}(x,p_T)&={\cal P}^q(\tilde p),\label{fperp:lfwf}\\
   x^2\,f^q_{4}(x,p_T)&=\frac{p_T^2+m_{q}^2}{2m_\pi^2}\,{\cal P}^q(\tilde p),
   \label{f4:lfwf}
\end{align}\end{subequations}
which formally coincide with the expressions for the unpolarized
nucleon TMDs \cite{Lorce:2014hxa}, except that the quark density operator 
${\cal P}^q(\tilde p)$ is evaluated in the pion states, which are
given in terms of the pion LFWFs by
   \begin{eqnarray}\label{phi-overlap2B} 
   {\cal P}^q(\tilde p)&=&\sum_{\lambda_1,\lambda_2} 
   | \Psi^{q \bar q}_{\lambda_1\lambda_{2}}( \tilde p ) |^2.
   \end{eqnarray} 
The expressions Eqs.~\eqref{f1:lfwf}-\eqref{f4:lfwf} are 
model-independent in the sense that they are valid in every 
light-front approach in which the Fock space expansion includes 
the leading (``valence'') sector, and truncates higher Fock space 
components.

\subsection{Internal consistency of the approach}
\label{Sec-3b:LFCM-consistency}

Let us now test the internal consistency of the approach. From 
Eqs.~\eqref{f1:lfwf}-\eqref{f4:lfwf} we obtain the relations
\begin{subequations}\begin{align}
x\,e^q(x,p_T)&=\frac{m_q}{m_\pi}\,f_{1}^q(x,p_T),\label{eq:e-freeq}\\
x\, f^{\perp q}(x,p_T)&=f_{1}^q(x,p_T),\label{eq:fperp-freeq}\\
x^2\,f_4(x,p_T) & =  \frac{p_T^2+m_q^2}{2m_\pi^{2}}\,f_1^q(x,p_T),\label{eq:f4-free}
\end{align}\end{subequations}
which coincide with the EOM relations \eqref{Eq:eom-e}-\eqref{Eq:eom-f4}, 
respectively, with 
vanishing tilde terms as expected for  free
on-shell partons described in terms of LFWFs.

The valence number sum rule \eqref{Eq:f1-sum-rule} and the momentum sum 
rule \eqref{Eq:mom-sum-rule} are satisfied in the LFCM by construction.
As a consequence of Eq.~\eqref{eq:fperp-freeq},  one finds 
 $\int{\rm d}x \, xf^{\perp q}(x)=N_q$ and
 $\sum_q\int{\rm d}x \, x^2 f^{\perp q}(x)=1$.

The sum rules for the first and second Mellin moment of $e^q(x)$
in Eqs.~\eqref{Eq:sigma-sum-rule} and~\eqref{Eq:Jaffe-Ji-sum-rule} are 
valid with the proofs analog to the nucleon case \cite{Lorce:2014hxa}.
The sum rule (\ref{Eq:Jaffe-Ji-sum-rule}) also follows directly
from Eq.~(\ref{eq:e-freeq}), which in addition implies a sum rule 
for the second moment $\sum_q\int{\rm d}x \, x^2 e^q(x)=m_q/m_\pi$.

The sum rule (\ref{Eq:f4-sum-rule}) for $f_4(x)$ is not supported 
in the LFCM of the pion, and also the qLIR (\ref{Eq:LIR-f4}) 
is not valid. These observations were also made in the nucleon case 
\cite{Lorce:2014hxa} and are related to each other. 
The fact that the same features occur in the pion (2-body) and nucleon
(3-body) case, indicates that this is not an artifact but a general 
property of  LFCMs. 
To ensure the compliance with the sum rule (\ref{Eq:f4-sum-rule}) it is 
necessary to consider zero modes in the light-front quantization 
\cite{Burkardt:1991hu} or to include higher light-front Fock states 
\cite{Brodsky:1998hn}. These considerations are beyond the scope of 
LFCMs based on the minimal Fock space, so that both the sum rule 
(\ref{Eq:f4-sum-rule}) and the qLIR (\ref{Eq:LIR-f4}) 
are consequently not satisfied \cite{Lorce:2014hxa}.
The LFCM of the pion complies, however, with positivity 
\eqref{Eq:f1-inequality}, \eqref{Eq:f4-inequality}.

Thus, the LFCM is internally consistent. It satisfies all general 
relations except for the sum rule (\ref{Eq:f4-sum-rule}) and the qLIR 
(\ref{Eq:LIR-f4}) which are beyond the scope of this approach, and 
both related to the academic twist-4 PDF $f_4^q(x)$ such that it
has no relevance for practical applications.

\subsection{Phenomenological model for LFWF}
\label{Sec-3c:LFCM-model-WF}

To obtain definite predictions one has to choose a specific model 
for LFWFs. In this work we choose the pion LFWFs proposed in 
Refs.~\cite{Schlumpf:1994bc,Chung:1988mu}.
One could include the effects of confinement in the light-cone
approach \cite{Brodsky:2006uqa}, but the phenomenological LFWFs of 
\cite{Schlumpf:1994bc,Chung:1988mu} provide already a phenomenologically 
acceptable description. They were
applied in Refs.~\cite{Pasquini:2014ppa,Frederico:2009fk} to the 
calculations of leading-twist $\mathsf T$-even and $\mathsf T$-odd TMDs, 
and generalized parton distributions of the pion. 
For completeness, we briefly review this model.

The explicit expression for the momentum-dependent part of the LFWF reads
\be\label{eq:psifc2}    
  	\tilde\psi_\pi(\bar x,{\bm \kappa}_{T }) 
   	= \sqrt{2(2\pi)^3}\;\sqrt{
	\frac{M_0(\bar x,{\bm \kappa}_T)}{4~\bar x (1-\bar x)}}\;
   	\frac{e^{-\bm{\kappa}^2/2\beta^2}}{\pi^{3/4}\beta^{3/2}}\,,
\ee
where $\bm{\kappa}=(\bm \kappa_T, \kappa_z)$ is the quark 
three-momentum, with
\begin{eqnarray}
	\kappa_{z}=M_0(\bar x,\bm{\kappa}_{T})\,(\bar x-\tfrac{1}{2}),
\end{eqnarray}   
and the free invariant mass squared is given by
\begin{eqnarray}
	M_0^2(\bar x,\bm{\kappa}_{T})=
	\frac{m^2_q+\kappa^2_{T}}{\bar x(1-\bar x)}.
\end{eqnarray}
The LFWF \eqref{eq:psifc2} depends on the free parameter $\beta$ and 
the quark mass $m_q$, which have been fitted to the pion charge radius 
and decay constant. In particular, we take $m_q=0.250$ GeV and 
$\beta=0.3194$~\cite{Schlumpf:1994bc}. For the spin-dependent part of 
the LFWF we refer to the derivation in Ref.~\cite{Pasquini:2014ppa}.
 
The results obtained with this pion LFWF model will be discussed,
and confronted with other models in Sec.~\ref{Sec-4:results}.

\section{Pion structure in bag and spectator model}
\label{Sec-4:other-models}

In this section we discuss pion TMDs in two other models, 
the bag and spectator model. We focus on physical aspects and 
internal consistency in these approaches, and skip technical
details which are collected in
\ref{App:details-bag} and \ref{App:details-spectator}.

\subsection{Bag model framework}
\label{Sec-3b:bag-model}

The bag model describes hadrons in terms of $n$ free quark 
and/or antiquark constituents confined inside a spherical cavity of 
radius $R_{\rm bag}$ by appropriate boundary conditions \cite{Chodos:1974je}.
In its simplest version $\pi$- and $\rho$-mesons are mass-degenerate, as it 
makes no difference whether a $\bar q q$-pair is placed in an $s$-wave 
with aligned or anti-aligned spins. This unrealistic situation can be 
improved \cite{DeGrand:1975cf} by invoking a gluon-exchange potential
	(which is an intrinsic property of the bag wave-function, 
	and different from the gluonic effects related to initial- 
	or final-state interactions \cite{Brodsky:2010vs}
	that give rise to $\mathsf T$-odd TMDs).	
Also ``center-of-mass corrections'' were used to construct wave-packet 
superpositions of static bag solutions with naturally light pion masses 
\cite{Donoghue:1979ax} that met phenomenological success \cite{Carlson:1982er}. 
A bag model version constructed to comply with chiral symmetry is the 
``cloudy bag'' \cite{Thomas:1982kv}.

In this work we use the simple MIT bag model
with massless quarks. At first glance this seems not to fit in
the generic picture of massive, effective, constituent degrees of freedom. 
But if desired, one can introduce a quark mass parameter 
with numerical but no conceptual differences in the model, 
with a value around $m_q \sim 120\,\rm MeV$ \cite{Golowich:1975vz} 
which is natural from the point of view of the constituent picture 
(although also smaller values were discussed in the literature).
More importantly, the quantum numbers of hadrons are determined 
by a fixed number of valence (quark, antiquark) degrees of freedom, which
allows one to classify the bag model as a constituent framework.
This approach is therefore sufficient
for our purposes to investigate generic features of TMDs in
constituent models. The bag model expressions for 
$f_1^q(x,p_T)$, $e^q(x,p_T)$, $f^{\perp q}(x,p_T)$, and $f_4^q(x,p_T)$ in the 
pion are given in \ref{App:details-bag-expressions}.

Keeping in mind the known general shortcomings, 
the description  has to be considered as consistent:
the bag model TMDs satisfy
the sum rules\footnote{
	\label{footnote-moments-in-bag-model}
	For that it is crucial to extend the integrals over the whole
	$x$-axis including negative $x$ and the regions $|x|>1$. 
	At negative $x$ the TMDs describe sea quarks where the bag results 
	violate positivity \eqref{Eq:f1-inequality},~\eqref{Eq:f4-inequality}
	in the nucleon case. The non-zero (albeit numerically small) 
	support in the regions $|x|>1$ is a technical problem in models 
	where corrections due to center-of-mass motion have to be applied. 
	These issues are known in the bag model of the nucleon 
	and not specific to the pion.}
\eqref{Eq:f1-sum-rule},~\eqref{Eq:mom-sum-rule},~\eqref{Eq:f4-sum-rule}.
The sum rules \eqref{Eq:sigma-sum-rule},~\eqref{Eq:Jaffe-Ji-sum-rule} 
are more subtle, and discussed in \ref{App:details-bag-sum-rules-e} 
where we show that they are  consistently satisfied in the model albeit in a quite different manner compared to QCD.
The bag results satisfy the inequalities 
\eqref{Eq:f1-inequality},~\eqref{Eq:f4-inequality}. As a last and
stringent consistency check of the description of higher-twist TMDs, 
we remark that the bag model satisfies the qLIR (\ref{Eq:LIR-f4}). 
This was proven analytically for nucleon TMDs in \cite{Lorce:2014hxa}. 
The proof can be carried over to the pion case 
such that also pion TMDs comply with Eq.~(\ref{Eq:LIR-f4}).
The EOM relations \eqref{Eq:eom-e}--\eqref{Eq:eom-f4}
hold with non-zero interaction-dependent tilde-terms 
which are due to bag boundary effects \cite{Jaffe:1991ra,Lorce:2014hxa}.

Overall we find that the bag model description of higher-twist TMDs
is internally consistent within the model, {\it although} not all 
features of the model are consistent with QCD. 
The PDFs in the bag model exhibit also interesting symmetry properties
which we discuss in detail in \ref{App:details-bag-symmetries-PDFs}.
We shall return to the bag model and discuss further properties of
TMDs and numerical results in section~\ref{Sec-4:results}.

\subsection{Spectator model}
\label{Sec-3c:spectator}

In the spectator approach the pion structure is modeled in terms of
an effective pion-quark-spectator 
vertex. The spectator has the quantum numbers of an antiquark but, 
constituting an effective degree of freedom,  it could in principle
have a different mass. We distinguish the spectator mass $M_R$
and constituent mass $m_q$ in the formulae 
in \ref{App:details-spectator}, 
but set them equal in the final results. 
This choice is closest to the spirit of constituent models where, 
after the active quark is struck, one would identify the ``remainder'' 
with an antiquark. This is of course not a necessary step. 
However the rational for working with a distinct effective 
degree of freedom is less convincing than in the nucleon case, 
where the ``remainder'' has the quantum numbers of diquarks, i.e.\ 
effective bosonic degrees of freedom whose masses are a priori
free parameters which {\it cannot} be associated with the constituent
quark mass. This approach was used to compute the pion TMDs $f_1^q(x,p_T)$, 
$f^{\perp q}(x,p_T)$, $e^q(x,p_T)$ in Ref.~\cite{Jakob:1997wg}.
In \ref{App:details-spectator-expressions} we review
the expressions for these TMDs, and derive also
the spectator model expression for $f_4^q(x,p_T)$.
 
Let us now concentrate on discussing the consistency of the approach 
which, regarding the sum rules \eqref{Eq:f1-sum-rule}-\eqref{Eq:f4-sum-rule},
is conceptually the same  in the spectator model of the pion as
in the spectator model of the nucleon \cite{Lorce:2014hxa}.
The valence sum rule (\ref{Eq:f1-sum-rule}) is satisfied in this model
by construction, as the normalization of the effective vertex is chosen 
adequately. In contrast, the momentum sum rule (\ref{Eq:mom-sum-rule}) 
is not valid for any choice of model parameters: one obtains less than 
unity in Eq.~(\ref{Eq:mom-sum-rule}). In a specific parametric limit,
one obtains a quasi model-independent result that the valence quark
and antiquark carry $\frac23$ of the pion's momentum. Such 
``$\frac23$-paradoxes'' have a long history in literature, and
illustrate that the model is incomplete, see the detailed
discussion in \ref{App:details-spectator-alpha-to-1+mom-sum-rule}.

The sum rules (\ref{Eq:sigma-sum-rule}) and (\ref{Eq:Jaffe-Ji-sum-rule}) 
for $e^q(x)$ do not hold in the spectator model of the pion.
This is apparent from the fact that the first and second moments in 
Eqs.~(\ref{Eq:sigma-sum-rule}) and (\ref{Eq:Jaffe-Ji-sum-rule})
should be positive, while $e^q(x)$ is negative 
in this model as discussed in \ref{App:details-spectator-fixing-alpha}.

Also the sum rule (\ref{Eq:f4-sum-rule}) for $f^q_4(x)$ is not satisfied 
in the spectator model, but this has a different origin. Both sum rules 
\eqref{Eq:f1-sum-rule} and \eqref{Eq:f4-sum-rule} can be traced back to the 
conservation of the Noether vector current. The form factors, which are 
introduced in an ad hoc manner to describe the effective vertex (see 
Eq.~(\ref{Eq:vertex-form-factor}) in \ref{App:details-spectator-expressions})
in general violate current conservation. It is therefore possible to satisfy 
(\ref{Eq:f1-sum-rule}) {\it or} (\ref{Eq:f4-sum-rule}) 
but not both sum rules simultaneously. 

The spectator model complies with the positivity requirement 
(\ref{Eq:f1-inequality}) for $f_1^q(x)$, and satisfies the inequality 
(\ref{Eq:f4-inequality}) for $f_4^q(x)$ 
provided one choses the model parameters appropriately, 
see \ref{App:details-spectator-fixing-alpha}.
As a last test of the spectator model, we notice that the qLIR
(\ref{Eq:LIR-f4}) is satisfied. The proof for that can be carried 
over from the nucleon case \cite{Lorce:2014hxa}. 

Finally, let us remark that the EOM relations 
\eqref{Eq:eom-e}-\eqref{Eq:eom-f4}
hold in the spectator model of the pion with the tilde-terms arising
due to the off-shellness of the quark, analog to the nucleon case 
\cite{Lorce:2014hxa}.
Remarkably, in the pion the off-shellness effects and hence 
the tilde-terms are large when one identifies the mass of the spectator
particle with the constituent quark mass.
This is discussed in detail in \ref{App-B}.

\section{Numerical results}
\label{Sec-4:results}

In order to discuss the model results, we first focus on the integrated TMDs 
in the three models in sections~\ref{subsec-4A:LFCM}--\ref{subsec-4C:spectator}.
Then we discuss the $p_T$-dependence of the TMDs in 
section~\ref{subsec-4D:pT-dependence}.

\begin{figure*}[t!]
   \begin{center}
   \epsfig{file=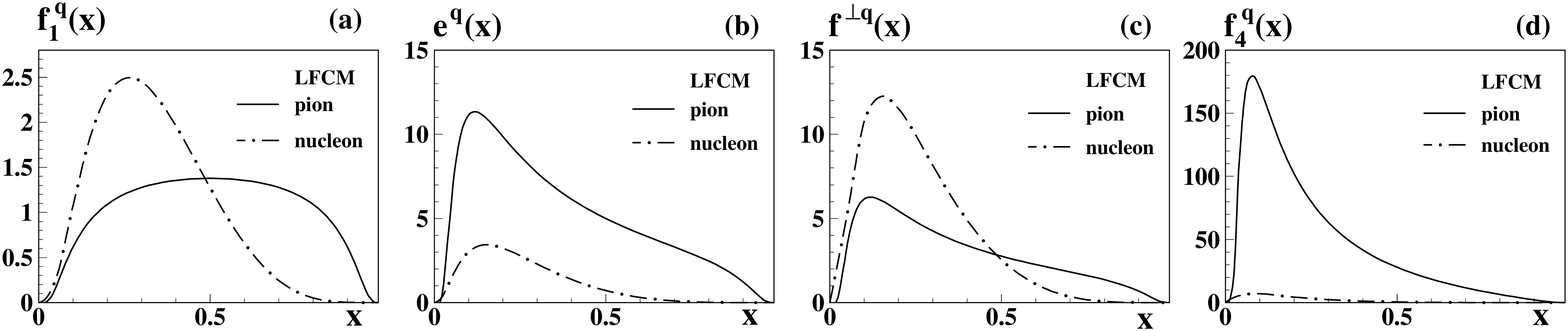,width=2\columnwidth}
   \end{center}
\vspace{-0.5 truecm}
   \caption{\label{fig1}
   	LFCM results for PDFs as functions of $x$. The solid curves 
	correspond to the pion results with the LFWF of 
	Ref.~\cite{Pasquini:2014ppa} for the $u$-flavor in $\pi^+$. 
	The dashed-dotted curves show for comparison the corresponding 
	results for the $d$-flavor PDFs in the proton in the 
	LFCM of Ref.~\cite{Lorce:2014hxa}, which have the same
	normalization for $f_1^q(x)$. }


   \begin{center}
   \epsfig{file=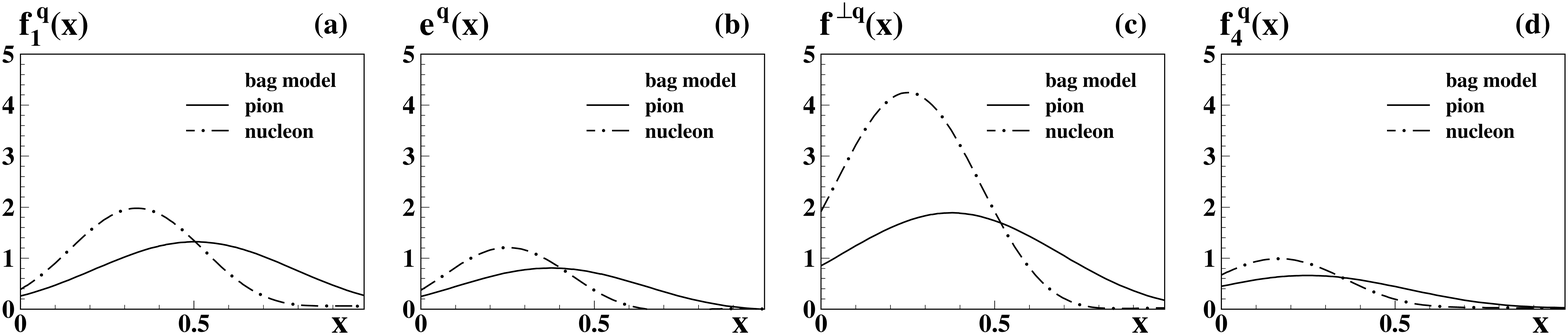,width=2\columnwidth}
   \end{center}
   \vspace{-0.5 truecm}
\caption{\label{FigXX:bag-model-results}
	Bag model results for pion PDFs (solid lines)
	as functions of $x$ at low scale:
	(a) $f_1^q(x)$,
	(b) $e^q(x)$,
	(c) $f^{\perp q}(x)$,
	(d) $f_4^q(x)$.
	The pion results (solid curves) refer e.g.\ to the $u$-flavor 
	in $\pi^+$. For comparison the corresponding nucleon PDFs from
	Ref.~\cite{Lorce:2014hxa} are shown (dashed-dotted curves) for 
	$d$-flavor in the proton, such that in panel (a) both curves 
	are normalized to unity
	(cf.\ footnote~3).
	}

  \begin{center}
 \epsfig{file=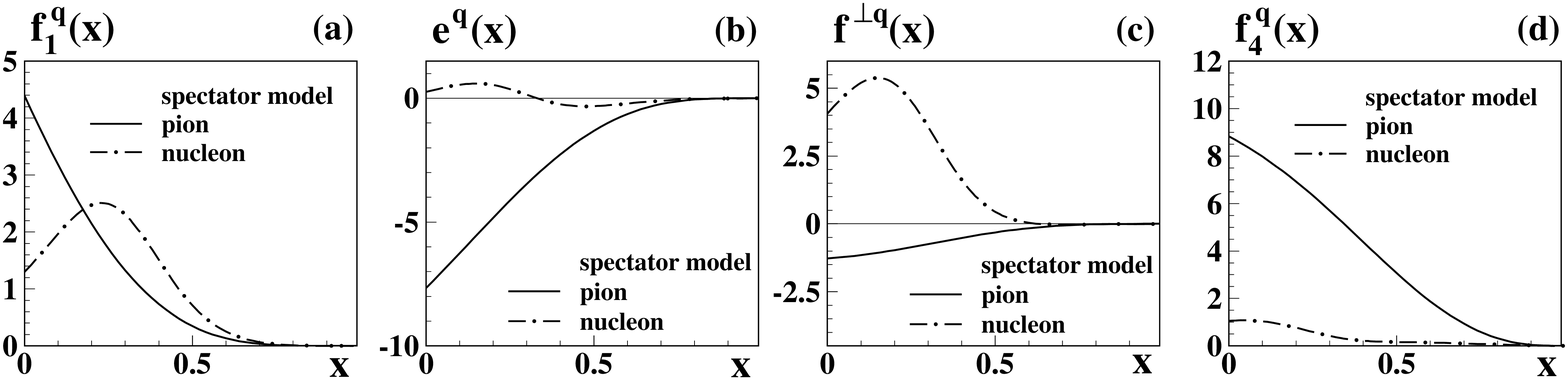,width=2\columnwidth}
  \end{center}
  \vspace{-4.5truecm}
  \caption{\label{fig3}
	Results for pion PDFs (solid lines) from 
	the spectator model as functions of $x$ at low scale:
	(a) $f_1^q(x)$,
	(b) $e^q(x)$,
	(c) $f^{\perp q}(x)$,
	(d) $f_4^q(x)$.
	The pion results  refer e.g.\ to the $u$-flavor in $\pi^+$.
	For comparison the corresponding nucleon integrated TMDs from 
	Ref.~\cite{Lorce:2014hxa} are shown (dashed-dotted curves) 
	for $d$-flavor in the proton, such that in panel (a) both 
	curves are normalized to unity.}

\end{figure*}

\subsection{Integrated TMDs in LFCM}
\label{subsec-4A:LFCM}

In figure~\ref{fig1} we show the LFCM results for the integrated TMDs 
$f_1^q(x)$, $e^q(x)$, $f^{\perp q}(x)$, and $f_4^q(x)$ of the pion in comparison 
with the corresponding results for the down quark in the nucleon, obtained 
from the three-quark LFWF of Refs.~\cite{Pasquini:2008ax,Lorce:2014hxa}.
In the LFCM the distribution of quark with longitudinal momentum fraction $x$ 
is equal to the distribution of the corresponding antiquark with longitudinal
momentum fraction $1-x$,
i.e.\ for instance in $\pi^+$ we have
$f_{1}^{u}(x)=f_{1}^{\bar d}(1-x)$.
Furthermore, one has the relation
$f_{1}^{\bar d}(x)=f_{1}^{u}(x)$
which gives as final result a momentum distribution symmetric with respect
to $x=\frac{1}{2}$.
The shape of the unpolarized momentum distributions for the 
pion and proton is quite different, reflecting the different valence-quark 
structure of the hadrons. For the proton, the unpolarized momentum 
distribution of the valence-quark is peaked at $x \approx 1/3$, 
while for the pion it reaches its maximum at $x=1/2$.

The twist-3 distributions of both the pion and the nucleon can be expressed 
in terms of the unpolarized momentum distribution as in 
Eqs.~\eqref{f1:lfwf}-\eqref{f4:lfwf}, with the corresponding hadron mass and 
constituent quark mass~\footnote{
	In the model calculation of Ref.~\cite{Pasquini:2014ppa} the mass of 
	the constituent quark in the nucleon was chosen as $m_q=0.263$ GeV.}.
The small value of the pion mass accounts for the enhancement of the $e^q$ 
and $f_4^q$ parton distributions with respect to $f^q_1$, which is much more 
pronounced than in the case of the nucleon, 
especially for $f_4^q$.

Finally, let us remark that in the LFCM it is possible to evaluate
also inverse moments. For instance, the inverse moment
\be\label{Eq:inverse-mom}
	\la x^{-1}\ra_q = \int_0^1\di x\;\frac{f_1^q(x)}{x}
\ee
exists and is well-defined in the LFCM. In fact, thanks to the 
EOM relations (\ref{Eq:eom-e}) and (\ref{Eq:eom-fperp}) it is related to the 
first moment of $f^{\perp q}(x)$ or the first moment of $e^q(x)$ (and by 
means of (\ref{Eq:sigma-sum-rule}) also to $\sigma_\pi$) in this model.
Such inverse moments have been discussed in the literature \cite{Brodsky:2007fr}
in the context of a modern reformulation of the Weisberger sum rule 
\cite{Weisberger:1972hk}. In general, in QCD as well as in the other 
models considered in this work, such inverse moments diverge and are 
ill-defined, so it is noteworthy that the LFCM provides a framework
where they can be evaluated --- giving the opportunity to study sum 
rules based on inverse moments. We will not pursue this line further 
in this work, and only remark that numerically one obtains
\be
	\la x^{-1}\ra_q = N_q\times\begin{cases} 
	2.82\, & \mbox{for pion (this work),} \\
	3.97\, & \mbox{for nucleon, Ref.~\cite{Lorce:2014hxa}.}
	\end{cases}
\ee

\subsection{Integrated TMDs in bag model}
\label{subsec-4B:bag}

The numerical results for the integrated pion TMDs from the bag model 
are shown in figure~\ref{FigXX:bag-model-results} in comparison to the 
results from the nucleon in this model \cite{Avakian:2010br,Lorce:2014hxa}.
For $f_1^q(x)$ the results are qualitatively similar in shape and magnitude 
to those from the LFCM.  But for 
$e^q(x)$, $f^{\perp q}(x)$, and $f_4^q(x)$ the
bag model predicts much smaller distributions than the LFCM.
This can be understood by means of the sum rules. In fact, $f_1^q(x)$ 
obeys the sum rules \eqref{Eq:f1-sum-rule} and \eqref{Eq:mom-sum-rule}
which dictate comparable magnitudes in all quark models.
On the other hand, the Jaffe-Ji sum rule (\ref{Eq:Jaffe-Ji-sum-rule})
does not place the same constraints regarding the magnitude of $e^q(x)$
in all models. The second moment of $e^q(x)$ is sizable in the LFCM
because the constituent mass 
$m_q=250$ MeV enters the normalization 
of this sum rule in the LFCM. In contrast to this, the quarks in the 
bag model are massless and the sum rule 
(\ref{Eq:Jaffe-Ji-sum-rule}) is realized differently,
see \ref{App:details-bag-sum-rules-e}, 
due to the different EOMs in the bag model.
Another principal difference is that the TMDs of the pion and
nucleon have the same order of magnitude in the bag model
in contrast to the LFCM.

There are several interesting observations, which we summarize
here leaving the details to \ref{App:details-bag-symmetries-PDFs}.
In the bag model $f_1^q(x)$ exhibits a global maximum at 
$x_{\rm max}\approx\frac1n$ where $n$ is the number of constituents,
and shows an approximate reflection symmetry 
$f_1^q(x) \approx f_1^q(2x_{\rm max}-x)$ which is satisfied numerically 
(for the pion with $n=2$) with an accuracy better than ${\cal O}(1\,\%)$ 
in the valence-$x$ region. As a consequence of this symmetry
the unpolarized distribution in the pion is smaller and broader than 
that in the nucleon, where $f_1^q(x)$ is approximately symmetric with 
respect to its peak at $x_{\rm max}\approx\frac13$.
These are natural features in a system made of $n$ constituents 
each one carrying 
on average about $x\sim\frac1n$ of the hadron momentum. 
With increasing $n$ one would expect the distributions 
to exhibit narrower peaks around their maxima, as we observe. 
We remark that $f_4^q(x)$ has similar properties to $f_1^q(x)$,
except that this PDF peaks at a different value 
$x_{\rm max}\approx \frac{1}{2n}$ 
and exhibits an approximate symmetry 
around this value, see \ref{App:details-bag-symmetries-PDFs}.

For pions the approximate symmetry $f_1^q(x) \approx f_1^q(1-x)$
implies that $f_1^q(x)$ has as much support at unphysical $x>1$ 
as in the region $x<0$ where it would describe {\it minus} the
distribution of antiquarks according to $f_1^{\bar q}(x)=-f_1^q(-x)$.
If we are willing to accept the spurious contributions at $x>1$
as a bag artifact (which can be remedied by adequate projection
techniques), then we recognize that the pion has no sea quarks
in the bag model, besides a spurious bag artifact contribution.
This is a qualitatively and quantitatively different situation
than in the nucleon, where $f_1^q(x)$ peaks around 
$x_{\rm max}\approx\frac13$ and the bag generates, through the 
symmetry $f_1^q(x) \approx f_1^q(\frac23-x)$,
sizable sea quark contributions in the nucleon which violate 
positivity (\ref{Eq:f1-inequality}).

With this last observation one arrives (somewhat paradoxically 
in view of the reservations regarding chiral symmetry) 
at the conclusion that the bag seems ``better suited'' for the 
description of the pion structure than the nucleon structure,
as the problem of unphysical sea quarks does not appear in
the pion case.

\subsection{Integrated TMDs in spectator model}
\label{subsec-4C:spectator}

In figure~\ref{fig3} we compare the integrated
TMDs $f_1^q(x)$, $e^q(x)$, $f^{\perp q}(x)$, and $f_4^q(x)$ 
from the pion spectator model 
with the parameter fixing as described in 
\ref{App:details-spectator-fixing-alpha} to the results in the
nucleon case obtained in \cite{Jakob:1997wg,Lorce:2014hxa}. 
Interestingly, and in contrast to other models and to
the nucleon case in the spectator model, the integrated
pion TMDs do not exhibit a global extremum at finite $x$,
but at the boundary value $x=0$.
The predictions for the functions $e^q(x)$ and $f^{\perp q}(x)$
of the pion and nucleon differ significantly in this model.
Although the description of these TMDs is conceptually the
same (one basically deals with the same effective diagram 
in the ``crossed channel'' \cite{Jakob:1997wg}), this is a 
consequence of the different parameters and the different
relative size of off-shellness effects in pion and nucleon,
see  \ref{App-B}.

\subsection{\boldmath $p_T$-dependence in models}
\label{subsec-4D:pT-dependence}

In this Section we turn our attention to the $p_T$-dependence of the TMDs. 
Let us define the mean transverse momenta $(n=1)$ and the mean squared 
transverse momenta $(n=2)$ in a generic TMD $j(x,p_T)$ as follows 
\be\label{Eq:define-mean-pT}
       \la p_{T}^n\ra= \frac{\int\di x\int\di^2p_T \,p_T^n\,j(x,p_T)}
                           {\int\di x\int\di^2p_T \,j(x,p_T)}.
\ee
If a TMD had exactly Gaussian $p_T$-dependence one would find for the ratio
\be\label{Eq:define-RG}
	R_G = \frac{2}{\sqrt{\pi}}\;\frac{\la p_T\ra}{\la p_T^2\ra^{1/2}}
\ee
the result $R_G=1$. This has been occasionally used as a
quick test to see to which extent a model supports Gaussian
$p_T$-behavior \cite{Boffi:2009sh} which is observed phenomenologically
in many DIS reactions \cite{D'Alesio:2007jt,Schweitzer:2010tt}. 
However, one should use
such tests with caution as the following results from the LFCM show.

\begin{figure*}[t!]
   \begin{center}
   \epsfig{file=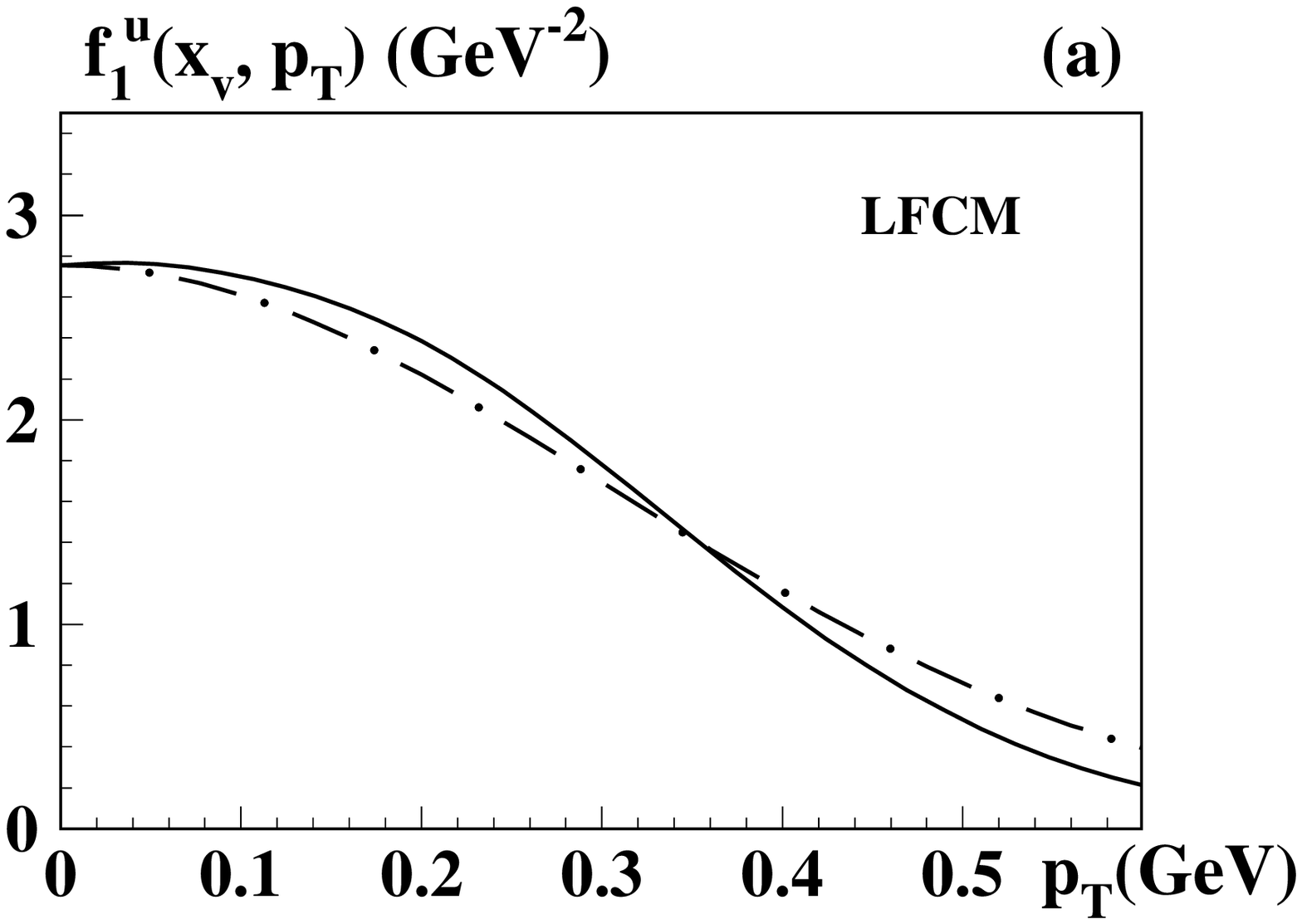,width=4.5 truecm}
	\hspace{-5mm}
   \epsfig{file=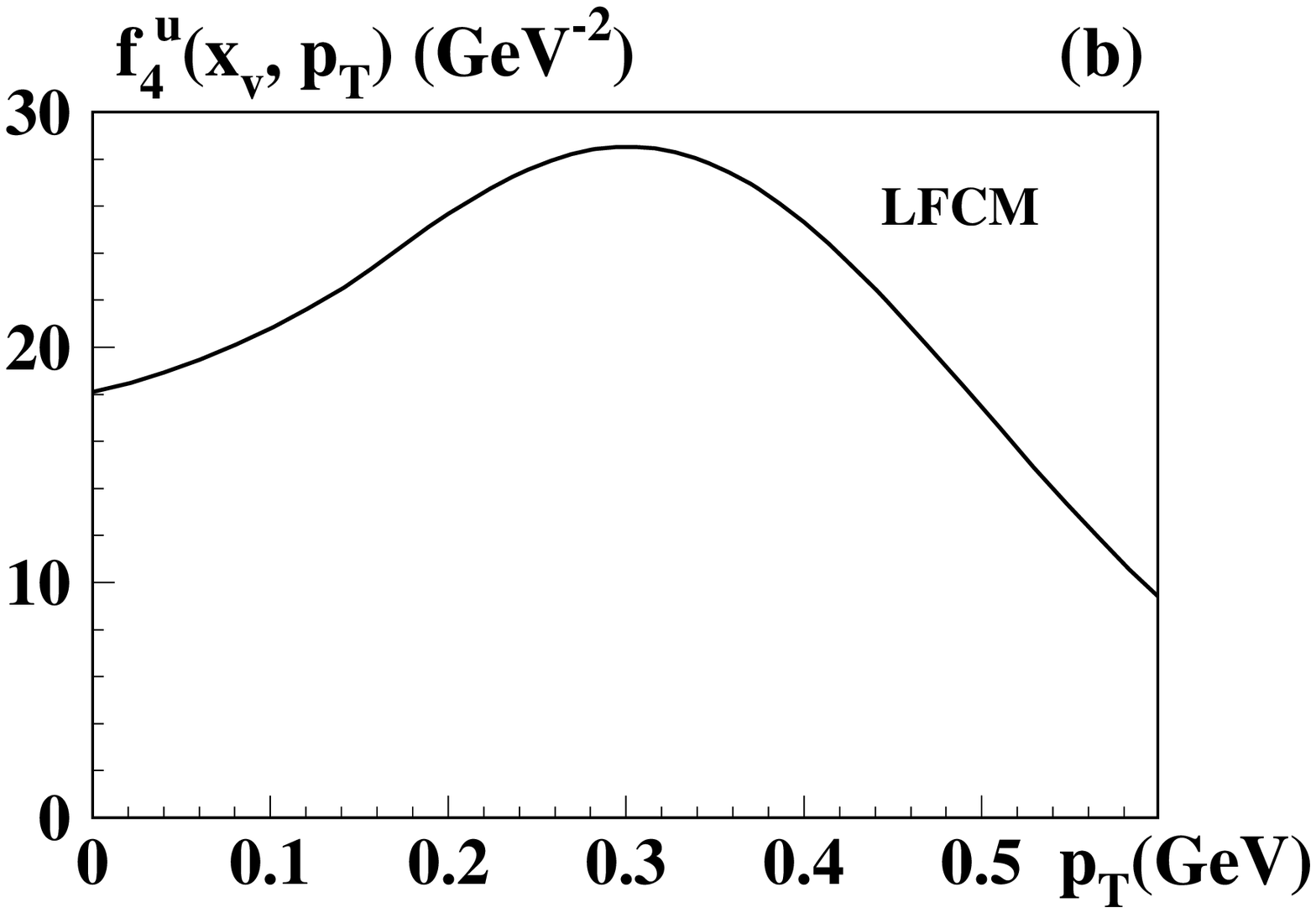,width=4.5 truecm}
   \end{center}
\vspace{-0.5 truecm}
   \caption{\label{fig4}
    (a) $f_1^u(x_v, p_T)$ 
    at $x_v=0.5$ as functions of $p_T$. The solid curves show 
    the predictions from the LFCM, while the dashed-dotted curves are 
    the respective Gaussian approximations from 
    Eq.~\eqref{gauss-model} 
    with the Gauss widths in Table~\ref{Table:pT-model-pion}$\,$(b). 
    (b) $f_4^u(x_v, p_T)$ 
    at $x_v=0.5$ as functions of $p_T$.
    We do not show results for $e^q(x,p_T)$ and 
    $f^{\perp q}(x,p_T)$ which differ merely in the overall normalization
    but exhibit the same $p_T$-dependence as $f_1^q(x,p_T)$.}
       \begin{center}
   \epsfig{file=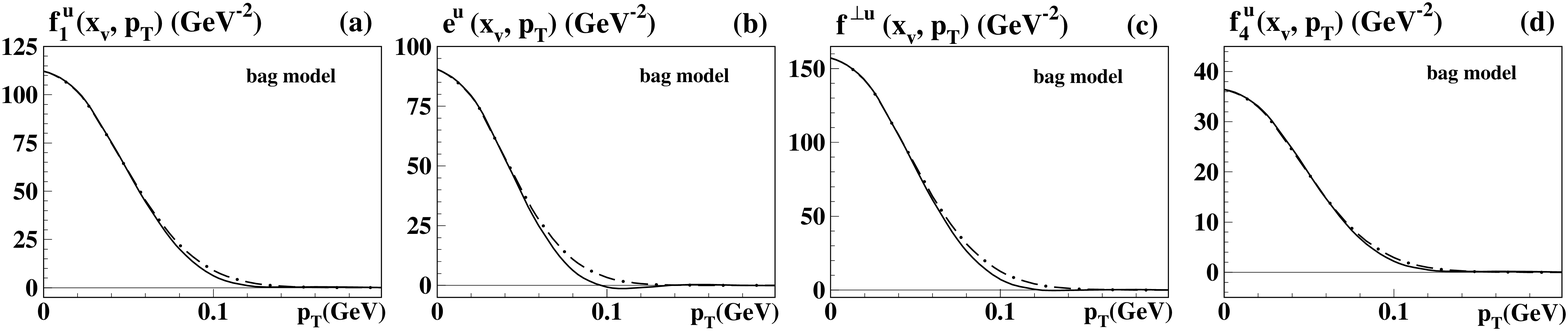,width=1.7\columnwidth}
   \end{center}
\vspace{-0.5 truecm}
   \caption{
   \label{FigXX:bag-model-tmd-results}
	Bag model results for pionTMDs at $x_v=0.5$ as functions of $p_T$
	 at low scale:
	(a) $f_1^q(x,p_T)$,
	(b) $e^q(x,p_T)$,
	(c) $f^{\perp q}(x,p_T)$,	
	(d) $f_4^q(x,p_T)$.
	The solid curves show the predictions from the bag model, while 
	the dashed-dotted curves are the respective Gaussian approximations 
	from Eq. \eqref{gauss-model} with the Gauss widths in 
	Table~\ref{Table:pT-model-pion}$\,$(b).}  
	   \begin{center}
\epsfig{file=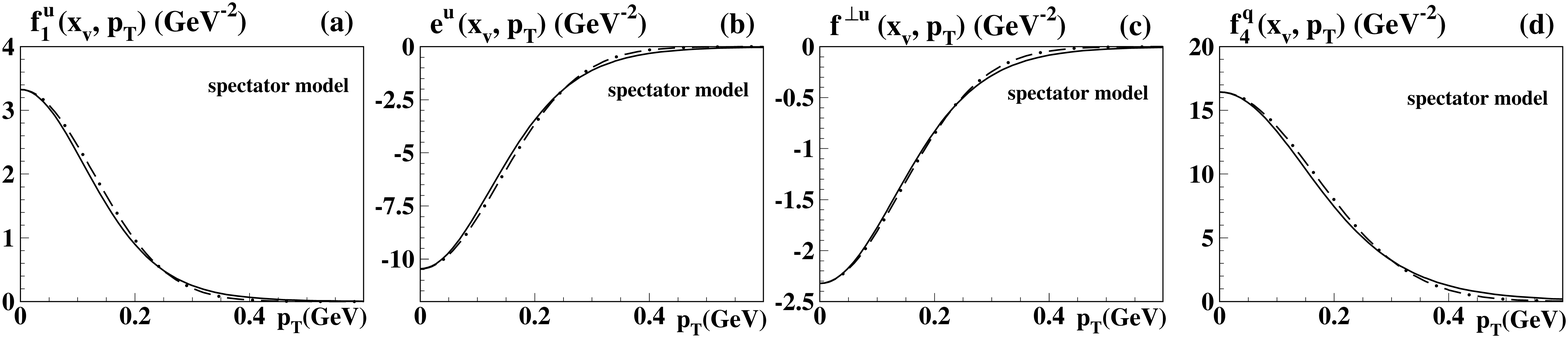,width=1.7\columnwidth}
   \end{center}
\vspace{-0.5 truecm}
   \caption{
   \label{FigXX:spectator-model-tmd-results}
	Spectator model results for pion TMDs at $x_v=0.5$ as functions of $p_T$
	 at low scale:
	(a) $f_1^q(x,p_T)$,
	(b) $e^q(x,p_T)$,
	(c) $f^{\perp q}(x,p_T)$,	
	(d) $f_4^q(x,p_T)$.
	The solid curves show the predictions from the spectator model for 
	$\alpha=3$  model, while the dashed-dotted curves are the respective 
	Gaussian approximations from Eq. \eqref{gauss-model} with the Gauss 
	widths in Table~\ref{Table:pT-model-pion}$\,$(b).}  
\end{figure*}

In Table~\ref{Table:pT-model-pion}$\,$(a) we show the results from the LFCM of
the pion for $\la p_T^{ }\ra$, $\la p_T^2\ra^{1/2}_{ }$ and the ratio $R_G$.
Although $R_G$ is very close to unity for all TMDs, the Gaussian Ansatz
is only a rough approximation for 
$f_1^q$, $e^q$, $f^{\perp q}$ and not applicable
at all for $f_4^q$, as shown in the right panel of figure~\ref{fig4}.

A more reliable 
test for the applicability of  the Gaussian Ansatz can be performed 
by introducing a different definition of 
$\la p_{T,v}^2\ra$~\cite{Avakian:2010br}, which is 
adjusted such that one obtains (if it is possible) a useful approximation 
of the true $p_T$-dependence of a TMD $j(x,p_T)$ at a given value of 
(valence-) $x$ in terms of the Gaussian Ansatz as 
\be\label{gauss-model}
     	j(x_v,p_T) \approx j(x_v,0) \; 
	\exp\biggl(-\frac{p_T^2}{\la p_{T,v}^2\ra}\biggr).
\ee
Although this definition is $x$-dependent, typically the $x$-dep\-endence 
is weak in the valence-$x$ region where quark models are applicable
\cite{Avakian:2010br}. For definiteness, we choose the value $x_v=0.5$ 
for the pion as a reference point where $f_1^q(x)$ exhibits a peak in most models.

In Table~\ref{Table:pT-model-pion}$\,$(b) the second column displays the results 
from LFCM of the pion for $\la p_{T,v}^2\ra^{1/2}$ of $f_1^q$, $e^q$, $f^{\perp q}$, where the Gaussian approximation is rough but still makes sense, see figure~\ref{fig4}. These 
numbers deviate significantly from the results for $\la p_T^2\ra^{1/2}$ in 
Table~\ref{Table:pT-model-pion}$\,$(a). 
The important lesson is that the ``$R_G$-test'' is only a necessary but not 
a sufficient condition for the usefulness of the Gaussian approximation.
Using the definition (\ref{gauss-model}), we can also directly compare all 
models, see the other columns in Table~\ref{Table:pT-model-pion}$\,$(b).
(Notice that the definitions (\ref{Eq:define-mean-pT}) would not
be useful in the bag model, where the integrations over $x$ 
in general include unphysical contributions, cf.\ 
footnote~\ref{footnote-moments-in-bag-model} 
and the discussion in section~(\ref{subsec-4B:bag})).
Comparing the models we see that the predictions for the widths
vary significantly from model to model. Notice that in the LFCM and
the spectator model the physical scale is set by the constituent quark 
mass, and the widths tend to be broader. In contrast to this, in the 
bag model the widths  $\la p_{T,v}^2\ra^{1/2}$ of the pion are substantially 
smaller. The reason is that the only dimensionful parameter in the bag model 
(here we work with massless ``current quarks'' confined in the bag) is the 
pion mass $m_\pi$ which is rather small.

Finally, for comparison we show in Table~\ref{Table:pT-model-nucleon}
the same information as in Table~\ref{Table:pT-model-pion}$\,$(b) but
for the nucleon in which case $x_v=0.3$ is a more appropriate choice
as this is where $f_1^q(x)$ peaks in quark models.
The nucleon results in Table~\ref{Table:pT-model-nucleon} are from 
Ref.~\cite{Lorce:2014hxa}\footnote{
	We would like to use this occasion to correct a numerical 
	mistake in the second column of Table~2 in 
	Ref.~\cite{Lorce:2014hxa}, where the widths in the LFCM of the 
	nucleon were incorrectly scaled by a factor of $1/\sqrt{\pi}$.
	The second column of Table~\ref{Table:pT-model-nucleon} in this work 
	gives the correctly scaled values. This correction does not affect 
	any of the conclusions of Ref.~\cite{Lorce:2014hxa}.}.

\begin{table*}
\begin{center}
\vspace{0.2 truecm}
\begin{tabular}{cc}
      \begin{tabular}{|c|c|c|c|}
      \hline
            {\bf pion}
      & \multicolumn{3}{c|}{ \ LFCM } \cr
      \hline
      TMD & \ $\la p_T\ra$ \ & \ $\la p_T^2\ra^{1/2}$ \ & \ $R_G$ \ \ \cr
      \hline
      $f_{1}^{q}$  & 0.28 &0.32 & 0.99 \cr 
      $e^{q}$     & 0.26 & 0.30 & 0.99 \cr
      $f^{\perp q}$ & 0.26 & 0.30 & 0.99 \cr
      $f_{4}^{  q}$ & 0.30 & 0.33 & 0.98 \cr
      \hline
            \end{tabular} \hspace{1mm} 
&      
      \begin{tabular}{|c|c|c|c|}
      \hline
      {\bf pion}
      & LFCM & bag & spectator \cr
      \hline
      TMD
      &  $\la p_{T,v}^2\ra^{1/2}$ 
      &  $\la p_{T,v}^2\ra^{1/2}$ 
      &  $\la p_{T,v}^2\ra^{1/2}$ \cr
      \hline
      $f_1^{q}$	 &  0.420 & 0.063 & 0.180 \cr
      $e^q$	 &  0.420 & 0.055 & 0.195 \cr
      $f^{\perp q}$&  0.420 & 0.063 & 0.200 \cr
      $f_4^{q}$	 &  --    & 0.063 & 0.235 \cr
      \hline
      \end{tabular} \hspace{1mm} 
\cr   
      {\bf\boldmath (a)} &  {\bf\boldmath (b)}
\end{tabular}\caption{\label{Table:pT-model-pion}
   	(a)
   	$\la p_T\ra$ and $\la p_T^2\ra^{1/2}$ in units of GeV as defined in 
	Eq.~\eqref{Eq:define-mean-pT}, and the ratio $R_G$ as defined in 
	(\ref{Eq:define-RG}) for pion TMDs  from LFCM.
     	(b)
   	The Gaussian widths $\la p_{T,v}^2\ra^{1/2}$ defined in 
	(\ref{gauss-model}) in GeV for pion TMDs at $x_v=0.5$
   	from LFCM, spectator and bag model. 	}
	\end{center}
\end{table*}

The comparison of the results for pion and nucleon 
in Table~\ref{Table:pT-model-nucleon} is very interesting.
We see that the three models make three different predictions.
In the LFCM the $p_T$-distri\-butions in the pion
are broader than those in the nucleon. In the bag model the
situation is opposite. In the spectator model the two hadrons have
comparable Gaussian widths.
Currently these predictions cannot be tested except for the case 
of $f_1^q(x,p_T)$, where phenomenological studies indicate that the 
$p_T$-distribution in $f_1^q(x,p_T)$ of the pion is broader than 
in the nucleon \cite{Schweitzer:2010tt}. This is in qualitative agreement 
with the predictions of the LFCM in Table~\ref{Table:pT-model-nucleon}.
One should keep in mind though, that the phenomenological result
was inferred from Drell-Yan data at center-of-mass energies of 
$\sqrt{s}\sim 23\,{\rm GeV}$ and refers to scales $Q > 4\,{\rm GeV}$ 
above the charmonium resonance region \cite{Schweitzer:2010tt}. 
\begin{table}[b!]
\begin{center}
\begin{tabular}{c}
      \begin{tabular}{|c|c|c|c|}
      \hline
      {\bf nucleon}
      & LFCM 
      & bag 
      & spectator $(u/d)$ \cr
      \hline
      TMD
      &  $\la p_{T,v}^2\ra^{1/2}$ 
      &  $\la p_{T,v}^2\ra^{1/2}$ 
      &  $\la p_{T,v}^2\ra^{1/2}$ \cr
      \hline
    $f_1^{q}$	& 0.240   & 0.280   & 0.200/0.270 \cr
    $e^q$   	& 0.240   & 0.230   & 0.160/0.180 \cr
    $f^{\perp q}$	& 0.240   & 0.270   & 0.180/0.230 \cr
    $f_4^{q}$  	& 0.350   & 0.170   & 0.180/0.250 \cr
      \hline
      \end{tabular}
\end{tabular}
\end{center}
\caption{\label{Table:pT-model-nucleon}
	For comparison, the same as Table~\ref{Table:pT-model-pion}$\,$(b) but for the nucleon and at
	$x_v=0.3$. Notice that in the LFCM of the nucleon and the bag model 
	the widths for $u$- and $d$-flavors are the same, 
	but not in the
	spectator model of the nucleon.
	}
	\end{table}
In contrast to this the LFCM results refer to a low scale 
$\mu_0 \sim 0.5\,{\rm GeV}$. For a more quantitative comparison
it is necessary to take carefully evolution effects into account.

\section{Conclusions}
\label{Sec-5:conclusions}

We have studied in constituent model frameworks the $\mathsf T$-even TMDs 
of the pion  focussing on higher twist, with the goal to establish common 
features, investigate the origins of tilde-terms, and compare the 
results to the description of unpolarized TMDs in the nucleon.
To avoid bias and minimize model dependence, we investigated several 
constituent models, including the LFCM, bag and spectator models.
The results give interesting insights on the internal structure of the 
pion in the valence-$x$ region. 

Our focus was on the aspects related to the modeling of 2-body dynamics 
of the $q\bar q$-pair in the pion as opposed to the 3-body 
dynamics in the nucleon state. 
The theoretical expressions and numerical results for all higher-twist pion 
TMDs $e^q$, $f^{\perp q}$, $f_4^q$ from the LFCM and bag model are new, 
and so are the spectator model expressions and results for $f_4^q$ 
(in that model expressions for $e^q$, $f^{\perp q}$ were quoted in
\cite{Jakob:1997wg} but numerical results have not been presented previously).

We addressed the question of how genuine QCD
inter\-action-dependent terms contribute to higher-twist TMDs and are 
modeled in constituent frameworks.
In LFCM the hadron states are obtained from a light-front Fock-space 
expansion in terms of free on-mass-shell parton states, with the essential 
QCD bound-state information encoded in the LFWF. Each constituent parton 
state obeys the free equation of motion.  Therefore, certain unintegrated 
relations among TMDs that are valid in free quark models are naturally 
supported in this approach for both the pion and nucleon case, but not all. 
In particular, relations involving the twist-4 unpolarized TMD $f^q_4$ 
are not satisfied for the pion, confirming the results obtained 
in the nucleon case. A fully consistent description of $f_4^q(x)$ in 
light-front formalism requires the inclusion of
zero modes or higher Fock states which go beyond the scope of the LFCM. 
Due to the academic character of the twist-4 function $f^q_4$
this is of no relevance for practical applications.

For comparison we  discussed results for pion TMDs in
bag and spectator model. We found that the 3 models make
different predictions especially for higher-twist TMDs. 
We also explored to which extent the approaches are compatible 
with a Gaussian shape of the transverse momentum distributions,
and found that all model results can be reasonably approximated
by a Gaussian $p_T$-shape, except for $f_4^q$ in the LFCM model.
In contrast to the bag model and the spectator model, 
the LFCM predicts broader $p_T$ distributions in the pion than in 
the nucleon, which is in qualitative agreement with phenomenology.
This may indicate that a more realistic description of the pion structure 
is achieved in the light-front approach than in the other models. More data
and phenomenological studied are needed to clarify the situation.

In the quark models discussed in this work, the pion was treated on the 
same footing as other hadrons, i.e. as a particle composed of the respective 
constituent degrees of freedom. It has to be regarded as a limitation 
that these models do not account for the nature of the pion as a 
Goldstone boson of chiral symmetry breaking. 
In view of the importance of chiral symmetry breaking, one may wonder to 
which extent we can trust the picture of the pion structure deduced from 
such models. We do not know the answer, but recently encouraging 
observations were made in this regard \cite{Lorce:2014hxa}.
In the nucleon chiral symmetry breaking effects were shown 
to have profound consequences for the sea quark structure, but
far less so for valence distributions \cite{Schweitzer:2012hh}. 
In fact, the description of valence quark distributions in
chiral models \cite{Schweitzer:2012hh} is qualitatively similar to 
those obtained in quark models \cite{Jakob:1997wg,Avakian:2010br}.
We are not aware of any argument why this situation should be
fundamentally different in the pion case, though it has not yet been 
investigated and remains an interesting question to address.
Another argument in favor of modeling pions and nucleons on the
``same footing'' in constituent approaches is based on the 
observation that pion and nucleon have similar sizes. In quark 
models like LFCM or spectator model, the scale for that is set 
by the constituent quark mass which also governs the $p_T$-behavior
of valence quark TMDs.
As a last encouraging observation, let us mention that in the LFCM
a phenomenologically rather successful description of the leading-twist 
pion structure (including the $\mathsf T$-odd Boer-Mulders function) was 
obtained \cite{Pasquini:2014ppa}. It of course remains to be tested in future
studies whether this success continues beyond leading twist.

Our results will provide useful guidelines for the interpretation of 
Drell-Yan data from pion-nucleon collisions, which are currently under 
study at the COMPASS experiment at CERN. These data are
expected to provide important insights on the (spin) structure of the 
nucleon. At the same time, these data will
provide the unique opportunity to gain valuable insights on the structure 
of the pion at both leading and subleading twist. In fact, both aspects
are tightly connected, and one can view it either way: the pion is used 
as a tool to investigate the spin structure of the nucleon, and polarized 
nucleons are used to shed new light on the structure of the pion.
In any case, a good understanding of the pion structure is indispensable
and worth exploring for its own sake.

\begin{acknowledgements}
This work was supported in part  by the National Science Foundation 
under Contract No.~1406298, the European Research Council (ERC) 
under the European Union's Horizon 2020 research and innovation programme 
(grant agreement No. 647981, 3D\-SPIN),
  and the Deutsche Forschungsgemeinschaft
   (grant VO 1049/1).\end{acknowledgements}

\appendix

\section{\boldmath Bag model in detail}
\label{App:details-bag}

In this Appendix we include the bag model expressions for pion TMDs and 
PDFs to make this work self-contained, then we discuss the sum rules 
for the twist-3 function $e^q(x)$, and investigate the symmetries of PDFs. 

\subsection{Expressions for TMDs in bag model}
\label{App:details-bag-expressions}

The bag model expressions for the pion TMDs,
\be\label{Eq:App-bag-def-TMD}
\begin{aligned}
  f_1^q(x,p_T) &= N_q\; A\,
  \biggl[t_0^2(p) +2\,\frac{p_z}{p}\,t_0(p) t_1(p) +t_1^2(p) \biggr], \\
  e^q(x,p_T) &= N_q\; A\,
  \biggl[t_0^2(p) -t_1^2(p) \biggr], \\
  f^{\perp q}(x,p_T) &= N_q\; A\,
  \biggl[2\,\frac{M_{\rm had}}{p}\,t_0(p) t_1(p) \biggr], \\
  f_4^q(x,p_T) &= N_q\,\frac{A}{2}
  \biggl[t_0^2(p) -2\,\frac{p_z}{p}\,t_0(p) t_1(p)  +t_1^2(p) \biggr]
\end{aligned}
\ee
with $p=\sqrt{p_z^2+p_T^2}$ and $p_z$ and hadron mass $M_{\rm had}$
given by
\be\label{Eq:App-bag-pz-Mh}
	p_z 
	    = \left(x-\frac{3}{4n}\right)M_{\rm had}\, ,
	\;\;\; M_{\rm had} = \frac43\,n\,\frac{\omega}{R_{\rm bag}} ,
\ee
coincide with those for unpolarized nucleon TMDs 
\cite{Lorce:2014hxa,Avakian:2010br}, if one considers the flavor structure, 
e.g.\ $N_u=N_{\bar{d}}=1$ for $\pi^+$, and writes the normalization constant 
$A$ in a way valid for mesons ($n=2)$ and baryons ($n=3$) as follows
\be\label{Eq:App-bag-norm}
	A=\frac{M_{\rm had}\omega^3R_0^3}{4\pi^2(\omega-1)\sin^2\omega}\;,
\ee
where $\omega \approx 2.04$ is the dimensionless ``frequency'' of the 
lowest bag eigenmode. In practice one uses the physical hadron mass for 
$M_{\rm had}$ and adjusts the bag radius accordingly. The functions 
$t_{l}(p)$ in Eq.~(\ref{Eq:App-bag-def-TMD}) can be expressed 
in terms of the spherical Bessel functions $j_l$ with $l=0,1$ as 
$t_{l}(p)=\int_0^1\ud u\,u^2j_l(upR_\text{bag})\,j_l(u\omega)$.

The bag model expression for $f_1^q(x,p_T)$ 
of the pion was also derived in \cite{Lu:2012hh}.

\subsection{Sum rules for $e^q(x)$ in bag model}
\label{App:details-bag-sum-rules-e}

The sum rules \eqref{Eq:sigma-sum-rule},~\eqref{Eq:Jaffe-Ji-sum-rule} 
can be evaluated analytically in the bag model,
and one finds for massless quarks
\begin{subequations}
\begin{align}
  \int \ud x\, e^q(x)   & =\frac{N_q}{2(\omega-1)}\;,
	\label{Eq:sum-rule-e1}\\
  \int \ud x\,x\,e^q(x)& =\frac{N_q}{2(\omega-1)}\,\frac3{4n}\;,
	\label{Eq:sum-rule-e2}
\end{align}
\end{subequations}
This is in contrast to QCD, where in the chiral limit  the
sum rule in Eq.~(\ref{Eq:sum-rule-e1}) should diverge
and the one in Eq.~(\ref{Eq:sum-rule-e2}) should vanish. 
These results reflect that the MIT bag model is at variance 
with chiral symmetry \cite{Jaffe:1991ra}. The problem can be 
traced back to the bag boundary condition, which breaks chiral symmetry. 
A trivial cure to this problem is to remove the bag boundary\footnote{
	A non-trivial cure to restore chiral symmetry is provided 
	by adequately ``matching'' chiral fields to the bag surface 
	as explored in the cloudy bag model of the nucleon
	\cite{Thomas:1982kv,Schreiber:1991tc}.
	}.
In this way one restores free quarks which comply with chiral symmetry 
in the massless limit. However, in this way one also removes the only 
interaction in this model, and all tilde-terms \cite{Lorce:2014hxa}. 
In particular, in this way $e^q(x,p_T)$ vanishes, as it is a pure 
bag-boundary effect \cite{Jaffe:1991ra}.
Interestingly, this in itself is consistent, because from the 
general decomposition in Eq.~(\ref{Eq:eom-e}) we see that 
in the chiral limit and in the absence of interactions one obtains 
$x\,e^q(x,p_T)=0$, although this does not necessarily imply that $e^q(x,p_T)$ 
itself must vanish as it could contain a $\delta(x)$-contri\-bution 
\cite{Efremov:2002qh,Schweitzer:2003uy}.
Notice however that Eq.~(\ref{Eq:sum-rule-e1}) is within the model
consistent, and provides the correct bag contribution to sigma-term 
as can be seen from the cloudy bag model study of the nucleon sigma 
term in Ref.~\cite{Jameson:1992ep}.

It is interesting to confront \eqref{Eq:sum-rule-e1},~\eqref{Eq:sum-rule-e2}
with the sum rules
$\int \di x\,f_1^q(x)=N_q$ and $\int \di x\,x\,f_1^q(x)=N_q/n$ showing 
that the bag model predicts that at low scale $e^q(x)$ is concentrated towards 
the region of lower $x$ as compared to $f_1^q(x)$,
\be
  \frac{\int\di x\,x\;e^q(x)}{\int\di x\,e^q(x)} = \frac34\;\frac1n\;\;\;
  {\rm vs.}\;\;
  \frac{\int\di x\,x\;f_1^q(x)}{\int\di x\,f_1^q(x)} = \frac1n\;.
\ee

\subsection{Symmetries of PDFs in bag model}
\label{App:details-bag-symmetries-PDFs}

From Eq.~(\ref{Eq:App-bag-def-TMD}) one obtains the following 
expressions for the PDFs (where $p_z$ retains its meaning as defined 
in Eq.~(\ref{Eq:App-bag-pz-Mh}), but $p$ denotes a dummy integration
variable in this section),
\be\label{Eq:App-bag-def-PDF}
\begin{aligned}
	f^q_1(x)	&=N_q\,2\pi\,A\int_{p_z}^\infty\ud p\;p
		\biggl[t^2_0(p)+2\,\frac{p_z}{p}\,t_0(p)t_1(p)+t_1^2(p)\biggr],\\
	e^q(x)		&=N_q\,2\pi\,A\int_{p_z}^\infty\ud p\;p
		\biggl[t^2_0(p)-t_1^2(p)\biggr],\\
	f^{\perp q}(x)	&=N_q\,2\pi\,A\int_{p_z}^\infty\ud p\;p
		\biggl[2\,\frac{M_{\rm had}}{p}\,t_0(p)t_1(p)\biggr],\\
	f^q_4(x)	&=N_q\,2\pi\,A\int_{p_z}^\infty\ud p\;p
		\biggl[t^2_0(p)-2\,\frac{p_z}{p}\,t_0(p)t_1(p)+t_1^2(p)\biggr].
\end{aligned}
\ee
To understand the exact and approximate symmetries of the PDFs in the
bag model, we need to recall
that $t_0(p)$ is an even function of $p$, while $t_1(p)$ is an odd
function of $p$. This implies that the integrands of all PDFs are
odd functions of $p$, i.e.\ in all cases the identity 
$\int_{-\kappa}^{\kappa} \dots = 0$ holds where the dots indicate the
respective integrands. If we choose $\kappa = p_z$ we immediately
conclude that for all PDFs one can equally well replace the lower 
integration limit by $(-p_z)$ or simply by $|p_z|$. This will be 
useful in the following.

The exact properties of $e^q(x)$ can be derived as follows. 
We can find the maximum of $e^q(x)$ by differentiating
\ba\nonumber
	\frac{\di}{\di x}\,e^q(x) = 
	-N_q\,2\pi\,A\,M_{\rm had}\;p_z\left[t^2_0(p_z)-t_1^2(p_z)\right]
	\stackrel{!}{=} 0	\\
\label{Eq:App-bag-derivativ-e}
	\;\;\;\;\Leftrightarrow\;\;\;\;
	{\rm(i)}\;\;
	p_z\stackrel{!}{=} 0 \;\; {\rm or} \; \;
	{\rm(ii)}\;\;t^2_0(p_z)\stackrel{!}{=}t_1^2(p_z)\;.
\ea
The condition (i) yields the position of the global maximum 
(as one can confirm by inspecting the second derivative)
\be\label{Eq:App-bag-xmax-exact}
	x_{\rm max}=\frac3{4n}\,.
\ee 
For completeness we remark that condition (ii) leads to many more
extrema with most of them appearing in unphysical regions of $x$.

Next, as we have seen above, the lower integration limit in 
Eq.~(\ref{Eq:App-bag-def-PDF}) can also be chosen as $|p_z|$. 
Since no factor of $p_z$ appears in its integrand, this means that 
$e^q(x)$ is a function symmetric under $p_z\mapsto -p_z$, i.e.\ it 
satisfies the exact symmetry
\be\label{Eq:App-bag-e-exact}
	e^q(2x_{\rm max}-x) = e^q(x)\;,\;\;\;x_{\rm max}=\frac3{4n}\,.
\ee
The above derivation can be repeated step by step with $f^{\perp q}(x)$.
Although it has a different shape, this PDF exhibits a global maximum
at the same position as $e^q(x)$ and satisfies also the same exact
symmetry
\be\label{Eq:App-bag-fperp-exact}
	f^{\perp q}(2x_{\rm max}-x) = f^{\perp q}(x)\;,\;\;\;x_{\rm max}=\frac3{4n}\,.
\ee

For $f_1^q(x)$ and $f_4^q(x)$ the situation is different, and no exact 
symmetry of the above kind exists due to the appearance of the explicit 
factor of $p_z$ in their integrands in Eq.~(\ref{Eq:App-bag-def-PDF}).
What one can derive in this case is the exact relation 
\be\label{Eq:App-bag-f1-f4-exact}
	f^q_1(x)=2f^q_4\biggl(\frac{3}{2n}-x\biggr),
\ee
though
the value $x=\frac{3}{2n}$ is not related to the maxima of $f^q_1(x)$ or 
$2f^q_4(x)$. One interesting application of Eq.~(\ref{Eq:App-bag-f1-f4-exact})
is that it immediately follows that $\int\di x\,f_1^q(x)=\int\di x\,2f_4^q(x)$ 
if one recalls the remarks in footnote~\ref{footnote-moments-in-bag-model}.

One can use the above method to find the maximum of $f_1^q(x)$.
The unpolarized function has its maximum at that value of $x$ 
which solves the integral equation
\be
	p_z \left(t_0(p_z)+t_1(p_z)\right)^2 = 2 
	\int_{p_z}^\infty \di p\;t_0(p)\,t_1(p)\, ,
\ee
where $x$ appears implicitly in $p_z$, see Eq.~(\ref{Eq:App-bag-pz-Mh}).
The solution can be found numerically and reads
\be
	x_{\rm max} = (1.00534\dots)\times\frac1n.
\ee
This is numerically very close to $x\approx \frac1n$ and an
intuitive result, see section~\ref{subsec-4B:bag}. 
There is also an approximate symmetry 
\be\label{Eq:approx-symm-f1}
	f_1^q(\frac2n-x) \approx f^q(x)
\ee 
which
for $n=2$ is satisfied to within ${\cal O}(1\,\%)$ 
accuracy for $x\in[0,1]$ (the approximate symmetry    
$f_1^q(2x_{\rm max}-x) \approx f_1^q(x)$ is much better in the vicinity
of $x_{\rm max}$ but interestingly overall somewhat worse).

The situation for $f_4^q(x)$ is similar to that of $f_1^q(x)$ 
except for the difference that the maximum appears at 
$x_{\rm max}\approx \frac1{2n}$. More precisely, for $f_4^q(x)$ 
one deals with the integral relation
\be
	p_z \left(t_0(p_z)-t_1(p_z)\right)^2 = -\,2 
	\int_{p_z}^\infty \di p\;t_0(p)\,t_1(p)\, ,
\ee
and the solution is
\be
	x_{\rm max} = (0.989327\dots)\times\frac1{2n}\,.
\ee

\begin{figure*}[t!]
   \begin{center}
    \epsfig{file=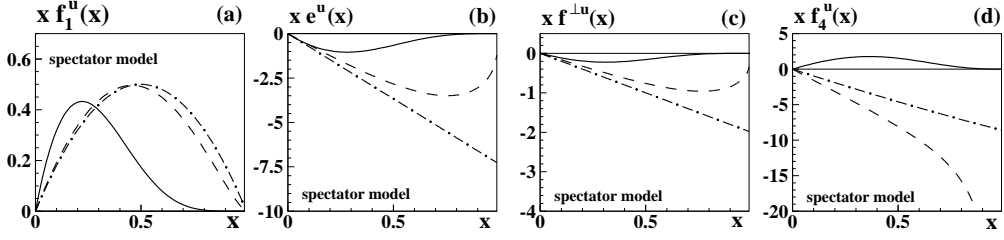,width=1.7\columnwidth}
   \end{center}
   \vspace{-4 truecm}
   \caption{\label{fig2}
	Integrated TMDs of up quarks in the pion from the calculation in the 
	spectator model of the pion~\cite{Jakob:1997wg} with $\alpha=1$ 
	(dashed-dotted curves),  $\alpha=1.2$ (dashed curves) and $\alpha=3$ 
	(solid curves).}

\end{figure*}

\section{\boldmath Spectator model in detail}
\label{App:details-spectator}

In this Appendix we review results for $f_1^q(x,p_T)$, 
$e^q(x,p_T)$ and $f^{\perp q}(x,p_T)$ from \cite{Jakob:1997wg}, 
derive in addition the expression for $f_4^q(x,p_T)$, and discuss 
the momentum sum rule, and parameter fixing in the spectator model.

\subsection{Expressions for TMDs in spectator model}
\label{App:details-spectator-expressions}

In the quark-spectator-antiquark model of the pion, the correlator 
(\ref{Eq:correlator-TMDs0}) is evaluated as follows
\begin{eqnarray}
  && \int \frac{\di z^-\di^2z_T}{2(2\pi)^3} \, 
   e^{i p \cdot z} \, \la P|\overline{\psi}(0)\Gamma \psi(z)|P\ra|_{z^+=0} \nonumber\\
	&&\equiv
   \left.\frac{\uTr[\tilde\Phi^q\Gamma]}{4(1-x)P^+}
   \right|_{p^2=xm^2_\pi-\tfrac{p^2_T+xM^2_R}{1-x}},
\end{eqnarray}
where
\be\label{correlator-spectator}
   \tilde\Phi^q=\frac{|g(p^2)|^2}{2(2\pi)^3}\,
   \frac{(\uslash\! p+m_q)(\,\uslash\! P+m_\pi)   (\uslash\! p+m_q)}{(p^2-m_q^2)^2},
\ee
with $g(p^2)$ a form factor. This form factor is often assumed to 
be~\cite{Melnitchouk:1993nk}
\be\label{Eq:vertex-form-factor}
   g(p^2)=N\,\frac{p^2-m^2_q}{|p^2-\Lambda^2|^\alpha},
\ee
where $\Lambda$ is a cut-off parameter and $N$ is a normalization constant. 
This choice has the advantage of killing the pole of the quark propagator.

The results for the quark TMDs of the pion  read
\be\label{Eq:TMDs-diquark}
\begin{aligned}
   f_1^{q}(x,p_T) &= B\,\frac{(m_q+xm_\pi)^2+p^2_T}{1-x},\\
   e^{q}(x,p_T) &= \frac{B}{(1-x)^2}\,
   \left[ (1-x)(m_q+xm_\pi)(m_q+m_\pi)\right.\\
     &\left.-M^2_R(x+\tfrac{m_q}{m_\pi})
 -(1+\tfrac{m_q}{m_\pi})p^2_T\right], \\
   f^{\perp q}(x,p_T) &= \frac{B}{(1-x)^2}\,
   \left[(1-x^2)m^2_\pi+2m_qm_\pi(1-x)\right.\\
   &\left.-M^2_R-p^2_T\right],\\
   f_4^{q}(x,p_T) &= \frac{B}{2(1-x)^2}\,
   \left\{(1-x)\left[(m_q+m_\pi)^2-M^2_R\right]\right.\\
   & \left.+\frac{p^2_T+M^2_R}{m^2_\pi}
     \left[\frac{p^2_T+M^2_R}{(1-x)}-2m_\pi m_q-(1+x)m^2_\pi\right]\right\}, 
\end{aligned}
\ee
where we introduced for convenience
\be
   B=\frac{N^2}{2(2\pi)^3}
   \left[\frac{1-x}{p^2_T+\lambda^2_R(x)}\right]^{2\alpha}\ee
    with
    \be
 \lambda^2_R(x)=(1-x)\Lambda^2+xM^2_R-x(1-x)m^2_\pi.
\ee

The results for the integrated TMDs read 
\begin{eqnarray}
 f_1^{q}(x)	&&=  B'\left[2(\alpha-1)(m_q+xm_\pi)^2+\lambda^2_R(x)\right],
		\nonumber\\
 e^{q}(x) 	&&=  \frac{B'}{1-x} \biggl\{2(\alpha-1)(x+\tfrac{m_q}{m_\pi})
		[(1-x)(m_q+m_\pi) m_\pi \nonumber\\
		&& -M^2_R]
		-(1+\tfrac{m_q}{m_\pi})\lambda^2_R(x)\biggr\},\nonumber\\
 f^{\perp q}(x)	 &&=  \frac{B'}{1-x}\biggl\{2(\alpha-1)[
		(1-x^2)m_\pi^2+2m_qm_\pi(1-x)\nonumber\\
		&& -M_R^2] -\lambda^2_R(x)\biggr\},\nonumber\\
  f_4^{q}(x) 	 &&=  \frac{B'}{2(2\alpha-3)m_\pi^2(1-x)^2}\,
		\biggl\{\lambda^4_R(x)- (2\alpha-3)\nonumber\\
		&& \times\lambda^2_R(x)
		\left[2m_qm_\pi(1-x)+(1-x^2)m_\pi^2-2M_R^2\right] \,\nonumber\\
		&&+ 2(\alpha-1)(2\alpha-3)\bigl[m_\pi^2(1-x)^2[(m_q+m_\pi)^2-M_R^2]\nonumber\\
		&&-M_R^2\left[2m_qm_\pi(1-x)+(1-x^2)m_\pi^2-M_R^2\right]\bigr] 
		\biggr\},\label{Eq:PDF-diquark}
\end{eqnarray}
where we introduced 
\be
   B'=\frac{N^2}{8(2\pi)^2(2\alpha-1)(\alpha-1)}
   \left[\frac{1-x}{\lambda^2_R(x)}\right]^{2\alpha-1}.
\ee

\

\subsection{Limit $\alpha\to 1$ and momentum sum rule}
\label{App:details-spectator-alpha-to-1+mom-sum-rule}

In the limit of $\alpha\to 1$  and $M_R\to m_q$
the PDFs reduce to, see Ref.~\cite{Jakob:1997wg},
\begin{subequations}\label{Eq:limit-alpha-1}
\begin{align}
   f_1^q(x)	& \stackrel{\alpha\to 1}{=} 2(1-x),
		  \label{Eq:f1-spec-limit-alpha-to-1} \\
   e^q(x)   	& \stackrel{\alpha\to 1}{=} -2
		  \left(1 +\frac{m_q}{m_\pi}\right),\\ 
   f^{\perp q}(x) 	& \stackrel{\alpha\to 1}{=} -2,\\
   f^q_4(x) 	& \stackrel{\alpha\to 1}{=} 
		\frac{2m_q^2-2m_qm_\pi(1-x)-m_\pi^2(1-x^2)
		-2\lambda^2_R(x)}{m^2_\pi(1-x)}.
\end{align}
\end{subequations}
The result (\ref{Eq:f1-spec-limit-alpha-to-1}) is interesting,
as it implies that $x\,f_1^q(x)$ is symmetric
under the exchange $x \leftrightarrow (1 - x)$.
But except for (\ref{Eq:f1-spec-limit-alpha-to-1}) the
results are unphysical, since the distributions do not vanish 
for $x\to 1$. Choices of $\alpha$ leading to acceptable results 
for all TMDs are discussed in \ref{App:details-spectator-fixing-alpha}.

Although the limit $\alpha\to 1$ in (\ref{Eq:limit-alpha-1}) is not
acceptable for all TMDs, the results ~(\ref{Eq:f1-spec-limit-alpha-to-1})
is useful for illustrative purposes. We shall work with this result to discuss 
the sum rules \eqref{Eq:f1-sum-rule} and~\eqref{Eq:mom-sum-rule}. 
The valence sum rule (\ref{Eq:f1-sum-rule}) is satisfied (in the limit
$\alpha\to 1$ and for $\alpha\neq 1$) though this is by construction, 
as the normalization constant $N$ is chosen adequately.
But the momentum sum rule (\ref{Eq:mom-sum-rule}) is not valid.
In the limit given by Eq.~(\ref{Eq:f1-spec-limit-alpha-to-1})
we obtain $\sum_q\int\di x\,x\,f_1^q(x)=\frac23$, where the sum goes 
over, e.g., the constituents $q=u,\,\bar d$ of the positive pion. 

Taken literally this result means the constituents carry only $\frac23$ 
of the hadron momentum. The deeper reason for this paradox can be traced 
back to the fact that the spectator model is an incomplete system as it 
does not account for the forces that would bind the constituents 
to form
a proper hadronic bound state 
which is essential\footnote{
	A situation of this type was encountered by Lorentz who found 
	$E=\frac23\,mc^2$ for the energy of an electron assumed to consist 
	of a charge distribution ``bound by some unknown forces of 
	non-electromagnetic origin'' \cite{Lorentz}.
	The latter citation is from Ref.~\cite{Diakonov:1996sr}, where 
	another $\frac23$-paradox of this nature occurs in a particular 
	approximation in the chiral quark-soliton model 
	which disappears when working in a fully consistent 
	solution of that model \cite{Weiss:1997rt}.}
to comply with the momentum sum rule \cite{Diakonov:1996sr}.
We note that $\sum_q\int\di x\, x\,f_1^q(x)<\frac23$ for $\alpha>1$.
Notice that in semi-phenomenological models based on the rainbow-ladder 
truncation of the QCD Dyson-Schwinger equations, one finds that valence quarks 
carry $\frac23$ of the pion momentum \cite{Chang:2014lva,Chen:2016sno}.
One could therefore be tempted to argue that the spectator model describes 
TMDs at somewhat higher scales, where valence quarks do not carry anymore 
$100\,\%$ of the hadron
momentum. However, this is phenomenologically 
not supported \cite{Jakob:1997wg} and must not distract 
from the fact that this model lacks the dynamics to form a consistent
bound state. 
\vspace{-0.45 truecm}

\subsection{Fixing of model parameters}
\label{App:details-spectator-fixing-alpha}

In the spectator model it is a priori not clear which value of $\alpha$ 
should be chosen in the form factors in Eq.~(\ref{Eq:vertex-form-factor}).
In figure~\ref{fig2} we therefore show the results from the spectator model 
of the pion for 
$xf_1^q(x)$, $xe^q(x)$, $xf^{\perp q}(x)$, and $xf_4^q(x)$
for different values of $\alpha$.
We  fix $M_R=m_q$, with $m_q=360$ MeV.

The dashed-dotted lines in figure~\ref{fig2} show results for $\alpha=1$ 
chosen in Ref.~\cite{Jakob:1997wg} 
for $f_1^q(x)$. This is not acceptable for the other TMDs 
which with this choice do not vanish for $x\to 1$, see  
\ref{App:details-spectator-alpha-to-1+mom-sum-rule}.
For $\alpha\ne 1$ the TMDs depend also on the cut-off $\Lambda$, 
that is taken equal to $0.4$ GeV as in Ref.~\cite{Jakob:1997wg}.
For $\alpha > 1$ one obtains $e^q(x)$ and $f^{\perp q}(x)$ which vanish
as $x\to 1$. To illustrate this point, we plot the results 
(dashed curves) for $\alpha=1.2$ in figure~\ref{fig2}. However, 
with this choice $f_4^q(x)$ is negative, violates the inequality
(\ref{Eq:f4-inequality}), and even diverges as $x\to 1$.
Both artifacts can be fixed by choosing $\alpha > \frac32$. The smallest
integer value $\alpha = 2$ would give a very large result for $f_4^q(x)$ 
with $\int\di x\,f_4^q(x) = 10.3$ strongly exceeding the sum rule 
(\ref{Eq:f4-sum-rule}). We plot therefore in figure~\ref{fig2} the results 
for $\alpha = 3$ (as solid lines) where $\int\di x\,f_4^q(x) = 3.6$ 
overestimates (\ref{Eq:f4-sum-rule}) less drastically (recall that the 
sum rule (\ref{Eq:f4-sum-rule}) cannot be satisfied for any $\alpha$).

We remark that one could vary the model parameters much more than that, 
e.g.\ one could vary the cutoff or relax the assumption that the spectator 
mass should be associated with the constituent quark mass $m_q$. But in this
work we shall content ourselves with the insight on the model dependence 
from the variation with respect to $\alpha$ in figure~\ref{fig2}.
The results shown in the main text were obtained for $\alpha=3$.

\begin{table}[h!]
\begin{center}

\begin{tabular}{c}

      \begin{tabular}{|c|c|c|c|}
      \hline
      	\ \textbf{(a) \ pion} \ & \multicolumn{3}{c|}{$p^2/m_q^2$}\\ 
	\hline
	\backslashbox{$\quad\, x$}{ $p_T\,$ (GeV)} 	& 0.1 & 0.2  &0.3	\\
		\hline
	0.1	& -0.18	& -0.44	&	-0.87\\
	0.2 	& -0.32	& -0.61	&	-1.09\\
	0.3 	& -0.49 & -0.82	&	-1.38\\
	\hline
	      \end{tabular} \\
       \ 	 \\
      \begin{tabular}{|c|c|c|c|}
      \hline
      	\ \textbf{(b) nucleon} \ & \multicolumn{3}{c|}{$p^2/m_q^2$ }\\ 
	\hline
	\backslashbox{$\quad\, x$}{ $p_T\,$ (GeV)} 	& 0.1 & 0.2  &0.3	\\
		\hline
	0.1	& 0.05	& -0.21	&	-0.64\\
	0.2 	& 0.03	& -0.26	&	-0.74\\
	0.3 	& -0.18    & -0.51	&	-1.06\\
	\hline
	      \end{tabular}	
      \end{tabular}
      \end{center}
	\caption{
	Off-shellness effects in the spectator model of pion (a)  and nucleon (b)
	at selected values of $x$ and $p_T$. If the active parton was onshell,
	the ratio $p^2/m_q^2$ would be unity and the tilde functions in
	Eqs.~\eqref{Eq:eom-e}--\eqref{Eq:eom-f4} would be absent.}\label{tab2}
\end{table}

\subsection{Off-shellness effects and tilde-terms}
\label{App-B}

The explicit expression for the tilde-terms in the spectator model with 
$M_R=m_q$ read
\be
\begin{aligned}
   x\,\tilde e^q(x,p_T)
   &=B\,\frac{p^2-m_q^2}{1-x}\,\left(x+\frac{m_q}{m_\pi}\right),\\
   x\,\tilde f^{\perp q}(x,p_T)
   &=B\,\frac{p^2-m_q^2}{1-x},\\
   x^2\,\tilde f^q_4(x,p_T)
   &=B\,\frac{p^2-m_q^2}{1-x}\,\frac{1}{2m^2_\pi}
   \biggl[[(m_q+xm_\pi)^2+p^2_T]\nonumber\\
   & \hspace{1cm}+x(1-x)m^2_\pi-\frac{x(p^2_T+m^2_q)}{1-x}\biggr],
\end{aligned}
\ee	
These terms arise from the off-shellness effects $p^2\neq m_q^2$.
We recall that in the spectator model the virtuality of the parton 
is given by $p^2 = xM^2_{\rm had}-(p^2_T+xM^2_R)/(1-x)$.
The results for the off-shellness effects at different values of $x$ 
and $p_T$  are shown in Table~\ref{tab2}(a) and \ref{tab2}(b) for the 
case of pion and nucleon, respectively. 
(The nucleon results are obtained with the axial diquark mass,
with the parameters used in \cite{Jakob:1997wg,Lorce:2014hxa}.)
We observe that these off-shellness effects are larger for the 
pion than for the nucleon, and the difference is more pronounced for 
small $x$ and moderate $p_T$.

In particular, the tilde-terms in $e^q(x)$ and $f^{\perp q}(x)$ are not
only sizable but also negative, and in fact overwhelm the contributions 
of the positive $f_1^q(x)$ in Eqs.~(\ref{Eq:eom-e}) and (\ref{Eq:eom-fperp}). 
This explains why $e^q(x)$ and $f^{\perp q}(x)$ are negative in the 
spectator model of the pion --- in contrast to the other models.
This feature is qualitatively different from the nucleon case,
where the tilde-terms could be viewed as ``corrections'' albeit
not necessarily small ones \cite{Lorce:2014hxa}. 
The reason for that is that in the pion case the constituent quark 
mass is larger than the hadron mass, i.e.\ off-shellness effects are
automatically more extreme than in the nucleon case. As a result, 
the spectator model of the pion does not support the Wandzura-Wilczek 
type approximation \cite{Avakian:2007mv} consisting in a neglect 
of tilde-terms.
\bibliographystyle{apsrev4-1}
\bibliography{biblio}

\end{document}